\documentclass{article}

\usepackage{graphicx, pdfpages}
\usepackage{natbib}
\usepackage{latexsym}
\usepackage{mathtools}
\usepackage{amsmath, amsthm, amssymb}

{\vskip 2pt \begin{compactitem}[#1]\setlength{\itemsep}{2pt}}
{\end{compactitem}\vskip 2pt}

\renewcommand{\citep}{\cite}
\renewcommand{\citet}{\cite}

\newcommand{\be}{\begin{equation}}
\newcommand{\ee}{\end{equation}} 
\newcommand{\lb}{\label}

\newcommand{\OL}{\overline}

\newcommand{\const}{({\rm const.})}

\newcommand{\bF}{{\bf f}}

\newcommand{\br}{{\bf r}}
\newcommand{\bu}{{\bf u}}

\newcommand{\bx}{{\bf x}}

\newcommand{\bU}{{\bf U}}
\newcommand{\bJ}{{\bf J}}
\newcommand{\bS}{{\bf S}}

\newcommand{\grad}{\bf \nabla}
\newcommand{\bdot}{\bf \cdot}
\newcommand{\bdots}{\bf :}

\begin{document}

\baselineskip=18pt
\begin{center}
\begin{LARGE}
{\bf Mapping the Energy Cascade in the North Atlantic Ocean: The Coarse-graining Approach}\\
\end{LARGE}

\bigskip
\bigskip
Hussein Aluie$^{1,2}$,
Matthew Hecht$^{3}$, and  Geoffrey K. Vallis$^{4}$\\
{\it
$^{1}$Department of Mechanical Engineering, University of Rochester, Rochester, NY 14627, USA,\\
$^{2}$Laboratory for Laser Energetics, University of Rochester, Rochester, NY 14627, USA,\\
$^{3}$Computational Physics and Methods (CCS-2), Los Alamos National Laboratory, Los Alamos, NM 87545, USA, \\
$^{4}$College of Engineering, Mathematics and Physical Science, University of Exeter University,  UK}

\bigskip
\bigskip

\begin{abstract}
A coarse-graining framework is implemented to analyze nonlinear processes, measure energy transfer rates and map out the energy pathways from simulated global ocean data. Traditional tools to measure the energy cascade from turbulence theory, such as spectral flux or spectral transfer rely on the assumption of statistical homogeneity, or at least a large separation between the scales of motion and the scales of statistical inhomogeneity. The coarse-graining framework allows for probing the fully nonlinear dynamics simultaneously in scale and in space, and is not restricted by those assumptions. This paper describes how the framework can be applied to ocean flows. 
\protect\\
Energy transfer between scales is not unique due to a gauge freedom. Here, it is argued that a 
Galilean invariant \emph{subfilter scale (SFS) flux} is a suitable quantity to properly measure energy scale-transfer in the Ocean. It is shown that the SFS definition can yield answers that are qualitatively different from traditional measures that conflate spatial transport with the scale-transfer of energy. The paper presents geographic maps of the energy scale-transfer that are both local in space and allow quasi-spectral, or scale-by-scale, dynamics to be diagnosed. Utilizing a strongly eddying simulation of flow in the North Atlantic Ocean, it is found that an upscale energy transfer does not hold everywhere. Indeed certain regions, near the Gulf Stream and in the Equatorial Counter Current have a marked downscale transfer. Nevertheless, on average an upscale transfer is a reasonable mean description of the extra-tropical energy scale-transfer over regions of $O(10^3)$ kilometers in size.
\end{abstract}
\end{center}

\vspace{0.5cm}

\clearpage

%

\section{Introduction}
\label{sec:intro}
Flow in the ocean is complex and very inhomogeneous, characterized by 
large-scale currents and a vast number of eddies. While much of the time-mean kinetic 
energy (KE) is concentrated in narrow intense currents such as the 
Gulf Stream and the Kuroshio Current, a substantial fraction of the total
KE is found at smaller scales in the time varying flow, largely at the 
mesoscale, where the size of eddies is established by the Earth's 
rotation and the ocean's stratification, with an important scale being 
the Rossby radius of deformation. The nature of the coupling between 
features spanning these scales, from the Rossby radius of deformation 
up to that of the large-scale mean flow, has long been of 
oceanographic interest. Our incomplete knowledge of the mechanisms 
that act to couple the mesoscale to the large scale circulation, and of 
the pathways through scales below the mesoscale by which energy is 
dissipated, has hindered our ability to fully account for the ocean's 
KE budget. There are additional reasons to engage in such study. From the perspective of
modeling, one must understand what processes are of fundamental
importance if those processes are liable to be compromised within the
model, as is often the case for processes involving mesoscale eddies
(e.g. \citep{Ringleretal13,Zannaetal17,Pearsonetal17}).

An enduring paradigm for oceanic energy pathways between large-scale and mesoscale flow \citep{Gilletal74,Rhines75,Salmon78,Salmon80,Smith_Vallis02,Vallis17,FerrariWunsch09}, is based on baroclinic instability and homogeneous quasigeostrophic (QG) turbulence theory. At large horizontal scales, there is a source of potential energy (PE), due to the wind and surface heat fluxes, that drives mesoscale eddies via baroclinic instability. The instability converts large-scale PE into KE at about the Rossby deformation scale of $R_d\approx 50-100$ km.  From this scale $R_d$, much of the KE, at least in this idealized model, undergoes some form of inverse cascade to larger scales. This paradigm is of course highly idealized whereas the World Ocean is irregular, highly inhomogeneous and constrained by topography and complex boundaries and, importantly, is not fully described by the QG equations. Even within the realm of QG dynamics, barotropic instabilities can arise to  transfer energy downscale.  One of the main objectives of this paper is to understand if and how this classical paradigm might apply in a more realistic situation, and as a first step we probe directly the KE transfer between scales in a comprehensive, strongly eddying ocean model. Specifically we analyze the energy transfer across scales at various geographic locations, such as in strong currents, near continental boundaries, and near the Equator.

Some intriguing and important work has already been done to examine the flow
of energy between different spatial scales in the oceans. For example, the work
of \citet{ScottWang05,Arbicetal13,Tullochetal11} represents largely
successful attempts to characterize turbulent scale-transfer as observed
from altimetry and generated within models, and the extent to which
those energy transfers conform to two dimensional geostrophic
turbulence. This and all previous oceanographic analyses, however,
have been generated using tools from turbulence theory that rely upon
an assumption of statistical homogeneity or, at least, a large scale separation between the eddying scales of motion and the scales over which the statistics vary. 

In this work, we try to relax this assumption by implementing a filtering approach that is mostly novel to large-scale physical oceanography but is well-established in other fluid dynamics
disciplines (e.g. \citep{Germano92,Meneveau94,Eyink95,Chenetal03,Aluie11,Riveraetal14}).
The approach is very general, mathematically exact, and based on 
a coarse-graining framework that can probe the dynamics of length-scales
at any geographic location and any instant of time, without relying on 
assumptions of homogeneity or isotropy. It can be used to analyze 
nonlinear processes, detect and measure energy transfer rates 
between oceanic structures, and map out energy pathways from 
ocean altimetry and model data. This paper presents an implementation of coarse-graining analysis 
for the quantification of oceanic energy flow across spatial scales. 
Our results indicate that the consequences of the assumption of statistical homogeneity embedded in
the traditional tools used for the analysis of turbulence can be
substantial, when applied in the context of oceanic flows. Whereas our
results are in many places in reasonable agreement with those from the
traditional method, they are different from the results of traditional
analyses in a number of energetic regions, indicating that the
assumption of homogeneity is, in those places, not justifiable.

Based on the evidence shown below, coarse-graining is found to 
be a viable method for exploring the degree to
which the generally accepted geostrophic model for such pathways is
valid in the ocean, and for studying the contribution of various
nonlinear mechanisms to the transfer of energy (or potential enstrophy) across scales, such as
baroclinic and barotropic instabilities, barotropization, Rossby wave
generation and internal wave generation and breaking. The method can also be applied to smaller scales where the geostrophic assumptions are not generally valid.

From the technical standpoint, this paper aims to introduce and prove 
the feasibility of the coarse-graining method in physical oceanography.
The hope is that it would enable the community to start 
mapping the energy pathways in the ocean, to identify the sources 
and sinks acting
at different scales, and to quantify the power rates at which they
generate or dissipate energy. The application of the method in this
paper is restricted to data from an eddy-resolving OGCM, thus allowing
us to probe the interaction of mesoscale eddies with the large
scales. However, we can also apply the method to simulations that
resolve submesoscale processes to probe the interaction of mesoscale
and submesoscale eddies with unbalanced motion, such as gravity waves,
and dissipative processes.  Indeed, the rather general applicability of the method can help lead to a
determination of the power requirements to sustain turbulence and
mixing, and the overall pathway of energy from source to sink,  in the ocean.

This paper is organized as follows. 
Section \ref{sec:method} introduces the coarse-graining
method in some detail and how we apply it to our data. Section 
\ref{sec:results} discusses the main results of this paper, and section
\ref{sec:PreviousStudies} offers a comparison of this work with previous
studies that have tackled these problems. The paper concludes with 
section \ref{sec:conclusion} which summarizes the main results and offers
ideas on potential future work and new research questions which we believe
the coarse-graining technique makes feasible.

\section{The coarse-graining method} \label{sec:method}

\begin{figure}
\centering
	\includegraphics[width=0.75\textwidth,height=0.3\textheight]{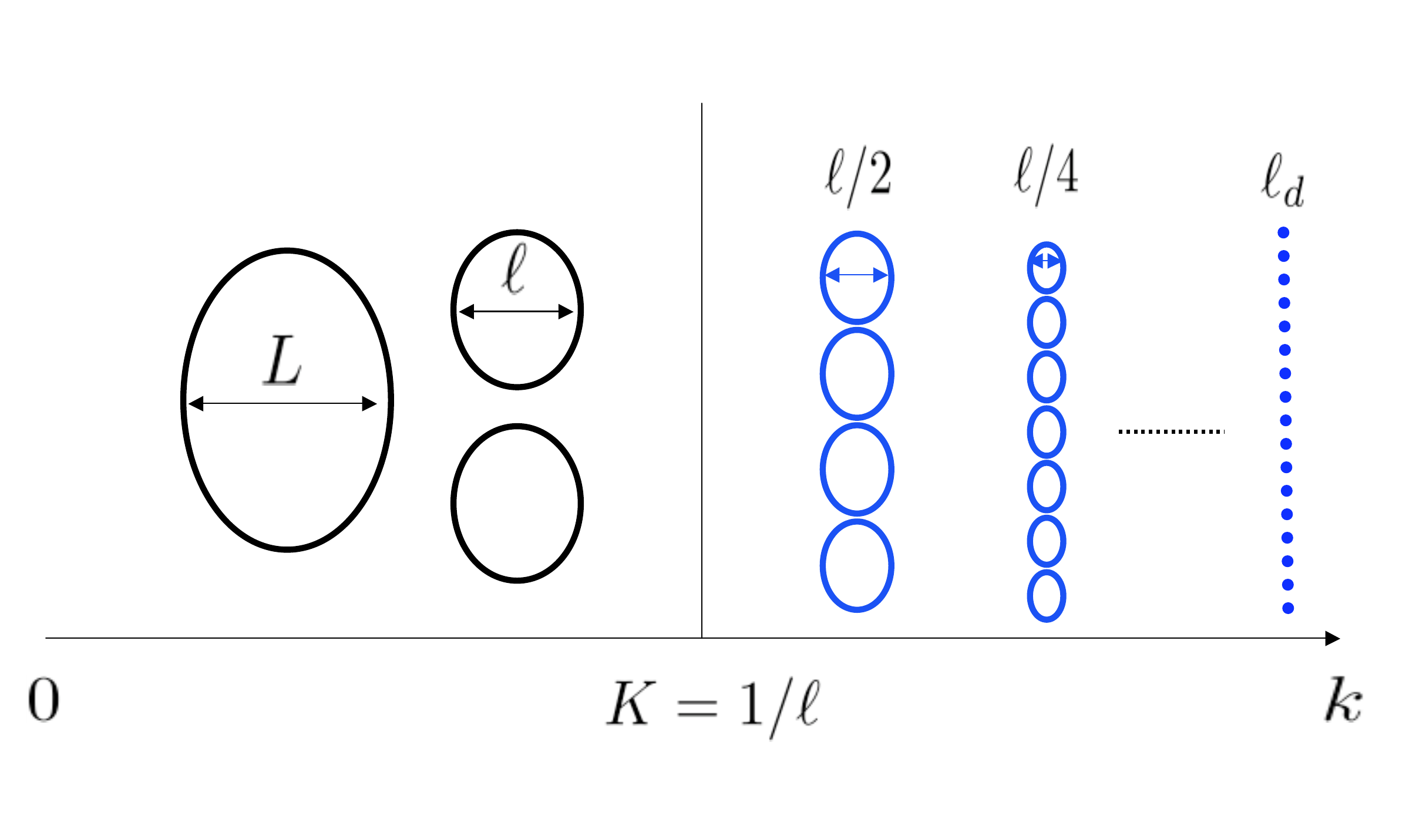}
	\caption{The coarse-graining approach. The system's size is $L$, the largest scale. Below the viscous dissipation scale $\ell_d$,  the dynamics is linear and modes are uncoupled. The dynamics over the entire scale-range $L\ge\ell\ge\ell_d$ is given from a numerical simulation or an experiment. Scales are then partitioned (post-processing) into large and small. Length $\ell$ represents the smallest scale that is resolved after coarse-graining. Scales $<\ell$ (in blue) are averaged out.}
	\label{fig:Filtering}
\end{figure}

In order to understand how energy travels through a system, both
geographically and with respect to scales (which we shall refer to as spatially and
spectrally, respectively) we use a ``coarse-graining'' or
``filtering'' framework that is unusual in large-scale physical oceanography but has 
become well-established in other fields. It is
rooted in a common technique in the mathematical analysis of partial
differential equations (e.g. \citep{Strichartz03,Evans10}).  It was first
introduced to the field of turbulence by \cite{Leonard74} in the
context of Large Eddy Simulation (LES) modeling. 
The method was further developed mathematically by 
\citet{Eyink95a,Eyink95b,Eyink05} to analyze the physics of scale 
coupling in turbulence. It has been utilized in several fluid dynamics 
applications, ranging from Direct Numerical Simulations (DNS) of turbulence
 (e.g. \citep{Piomellietal91,Vremanetal94,AluieEyink09}), to 
 2D laboratory flows in a shallow tank (e.g. \citep{Chenetal06,KelleyOuellette11,LiaoOuellette15,FangOuellette16})
 and in soap films (e.g. \citep{Riveraetal03,Chenetal03,Riveraetal14}), 
 to experiments of turbulent jets \citep{Liuetal94} and flows through a grid \citep{Meneveau94}, through a duct \citep{Taoetal02}, in a water channel \citep{Baietal13}, and in turbomachinery (e.g. \citep{Chowetal05,AkbariMontazerin13}). Moreover, the framework has been extended to rotating stratified flows
 \citep{AluieKurien11}, magnetohydrodynamics \citep{Aluie17b}, and compressible turbulence (e.g. \citep{AluieLiLi12}), and as a framework for parameterizing convection \citep{Thuburnetal17}.
The schematic in Figure \ref{fig:Filtering} summarizes the main idea behind the method.

The technique allows for a direct quantification of the strong (or
weak) nonlinear coupling between different scales. For example, it
allows one to measure the amount and sense (upscale or downscale) of energy being
exchanged between different scales at every point $\bx$ in the domain,
at every instant in time $t$. It is a very general approach to
analyzing complex flows, the rigorous foundation of which was
developed by \citet{Germano92,Eyink95,Eyink05} to analyze the
fundamental physics of scale interactions in turbulence. The method
allows for probing the dynamics simultaneously in scale and in space,
and is not restricted by usual assumptions of homogeneity or isotropy.
This makes it ideally suited for studying, on the entire globe, oceanic 
flows with complex continental boundaries. We have recently developed 
and generalized the approach to account for the spherical geometry 
of the flow \citep{Aluie17a}, with this work being its first implementation in a realistic 
geophysical system.

\subsection{Coarse-grained fields\lb{sec:FilteringScalars}}
The essence of the method is relatively straightforward. For any scalar field $f(\bx)$,
a ``coarse-grained'' or (low-pass) filtered field, which contains modes
at length-scales $>\ell$, is defined as
\be
\OL f_\ell(\bx) = G_\ell * f,
\lb{def:filtering}\ee
where $*$ is a convolution and $G_\ell(\br)$ is a normalized kernel (or window function) so that $\int d^2\br ~G_\ell(\br)=1$. 
Operation (\ref{def:filtering}) may be interpreted as a local space average over a region of diameter $\ell$ centered at point $\bx$. 
Notice that $\OL{f}_\ell (\bx)$ has scale information $\ell$ as well as space information $\bx$. 
An example of a kernel $G_\ell$ is the Top-hat kernel,
\begin{eqnarray}
H_\ell(\br)&=&\begin{cases}
    A^{-1}, & \text{if $|\br|<\ell/2$}.\\
    0, & \text{otherwise}.\\
  \end{cases}\lb{app_eq:Tophat_x}
\end{eqnarray}
In a flat (Euclidean) 2D domain, the normalization area is
$A=(\pi\ell^2)/4$, whereas on Earth's spherical surface, $A = 2\pi R^2
\left[1-\cos\left(\ell/2R \right) \right]$, where $R$ is Earth's
radius. It might be possible to use more general anisotropic 
kernels to distinguish between zonal and meridional scales, for example. 
For simplicity, we restrict ourselves to isotropic kernels in this paper and 
defer such refined analysis to future work.  Moreover, while the filtering can be done in all 
three dimensions, here we focus on the analysis of horizontal scales and filter
using 2D kernels to study the scale-transfer.

We can also define a complementary high-pass filter which retains only modes
at scales $<\ell$ by
\be  f^{'}_\ell(\bx) = f(\bx)-\OL{f}_\ell(\bx),
\lb{def:high-pass}\ee
which also retains spatial information as a function of $\bx$ and
scale information as a function of $\ell$.  In the rest of our paper, 
we shall omit subscript $\ell$ whenever there is no risk of ambiguity.

\begin{figure}
\vspace{-1cm}
\centering
	\includegraphics[trim=1cm 1cm 1cm 1cm,clip, width=0.49\textwidth,height=0.3\textheight]{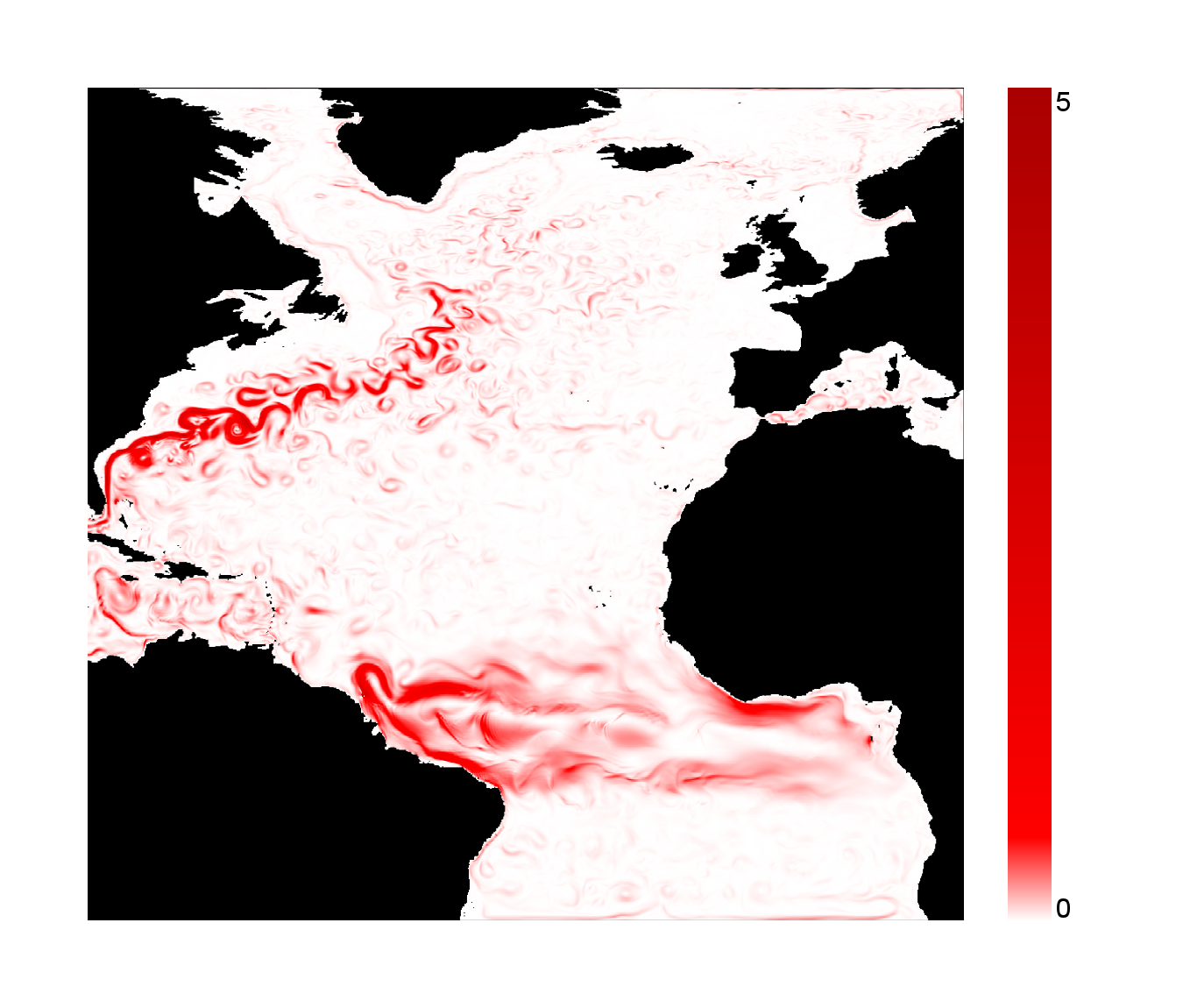}
\vspace{-0.5cm}	
	\includegraphics[trim=1cm 1cm 1cm 1cm,clip, width=0.49\textwidth,height=0.3\textheight]{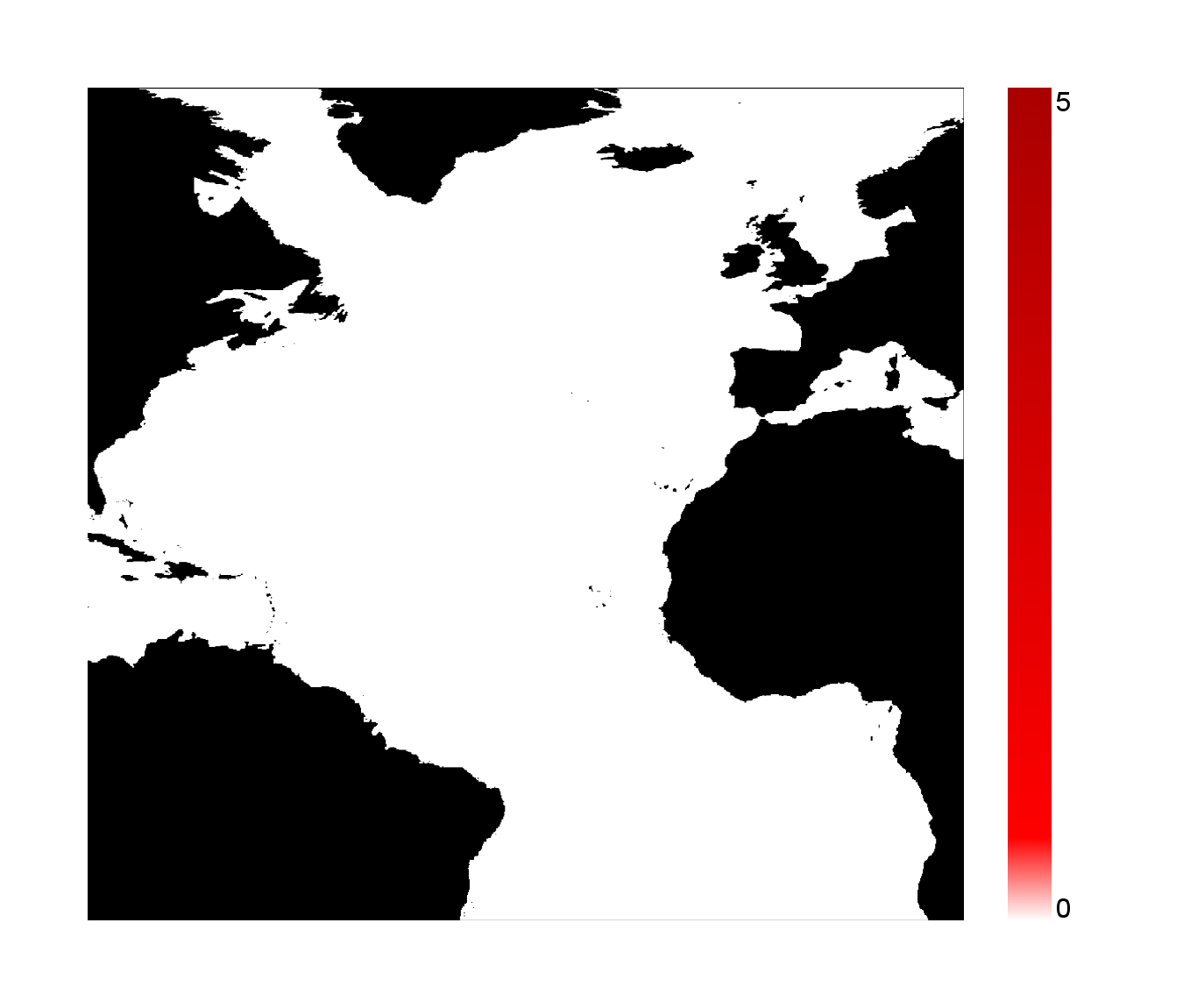}
\vspace{-0.5cm}	
	\includegraphics[trim=1cm 1cm 1cm 1cm,clip, width=0.49\textwidth,height=0.3\textheight]{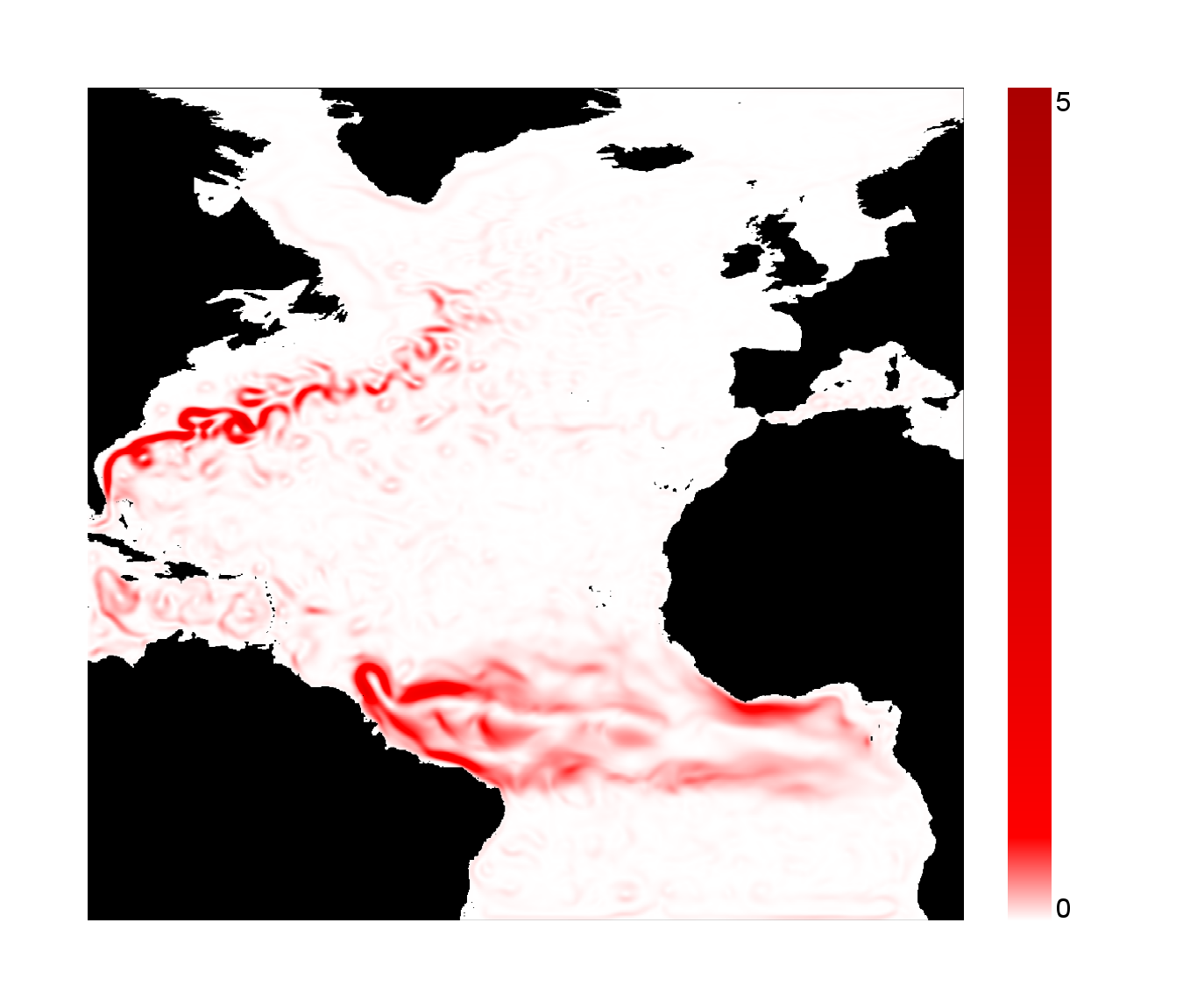}
	\includegraphics[trim=1cm 1cm 1cm 1cm,clip, width=0.49\textwidth,height=0.3\textheight]{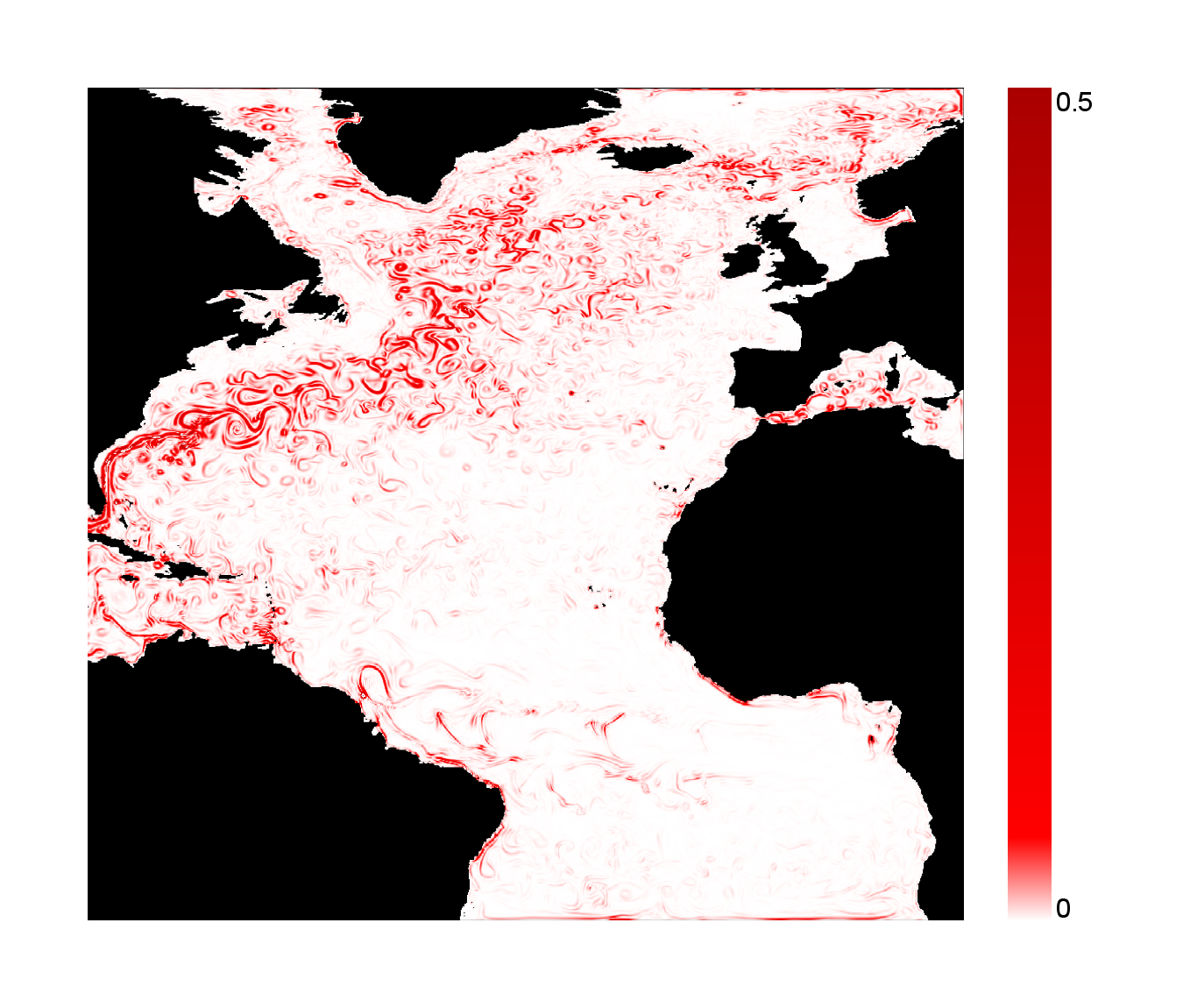}
	\includegraphics[trim=1cm 1cm 1cm 1cm,clip, width=0.49\textwidth,height=0.3\textheight]{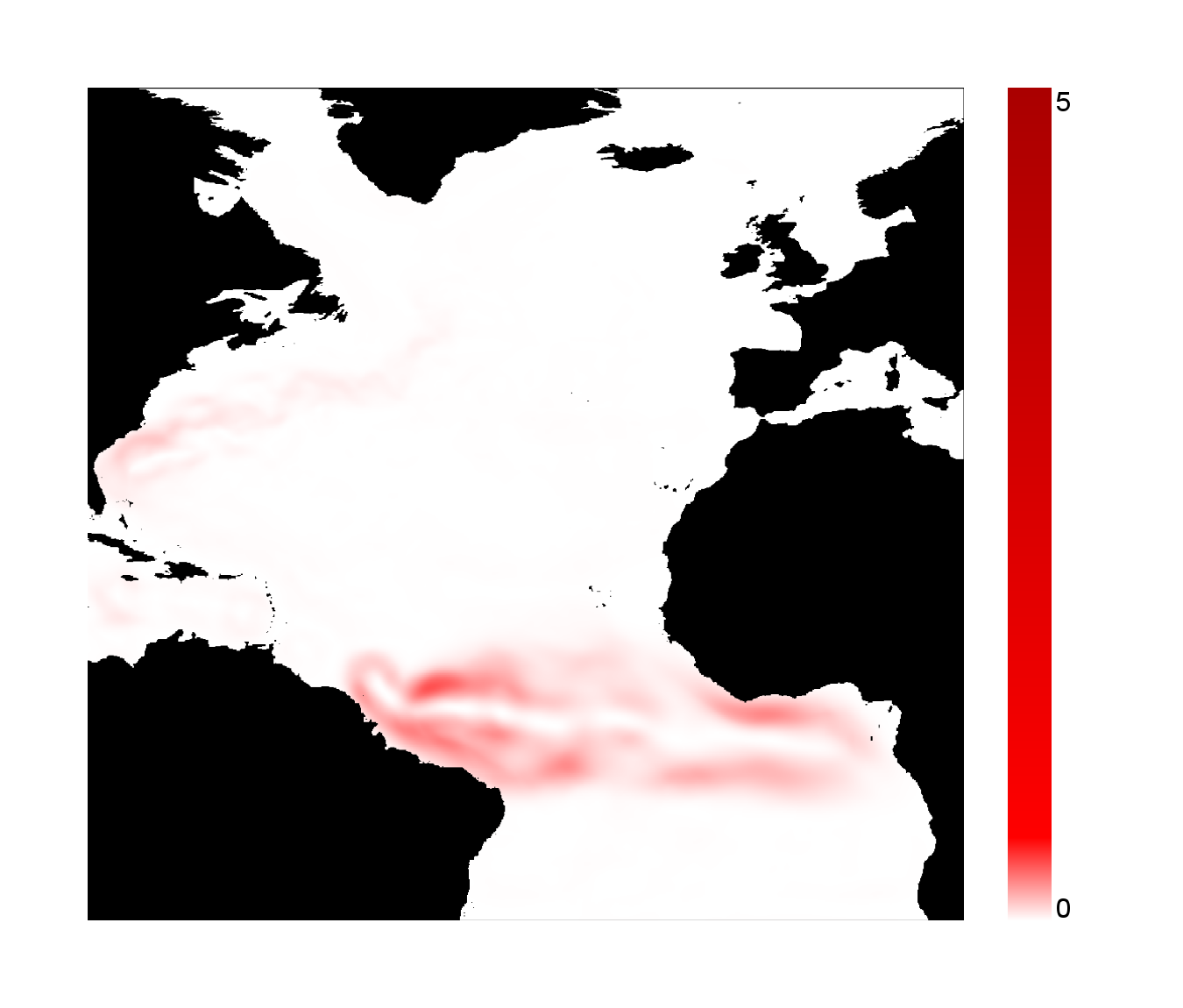}
	\includegraphics[trim=1cm 1cm 1cm 1cm,clip, width=0.49\textwidth,height=0.3\textheight]{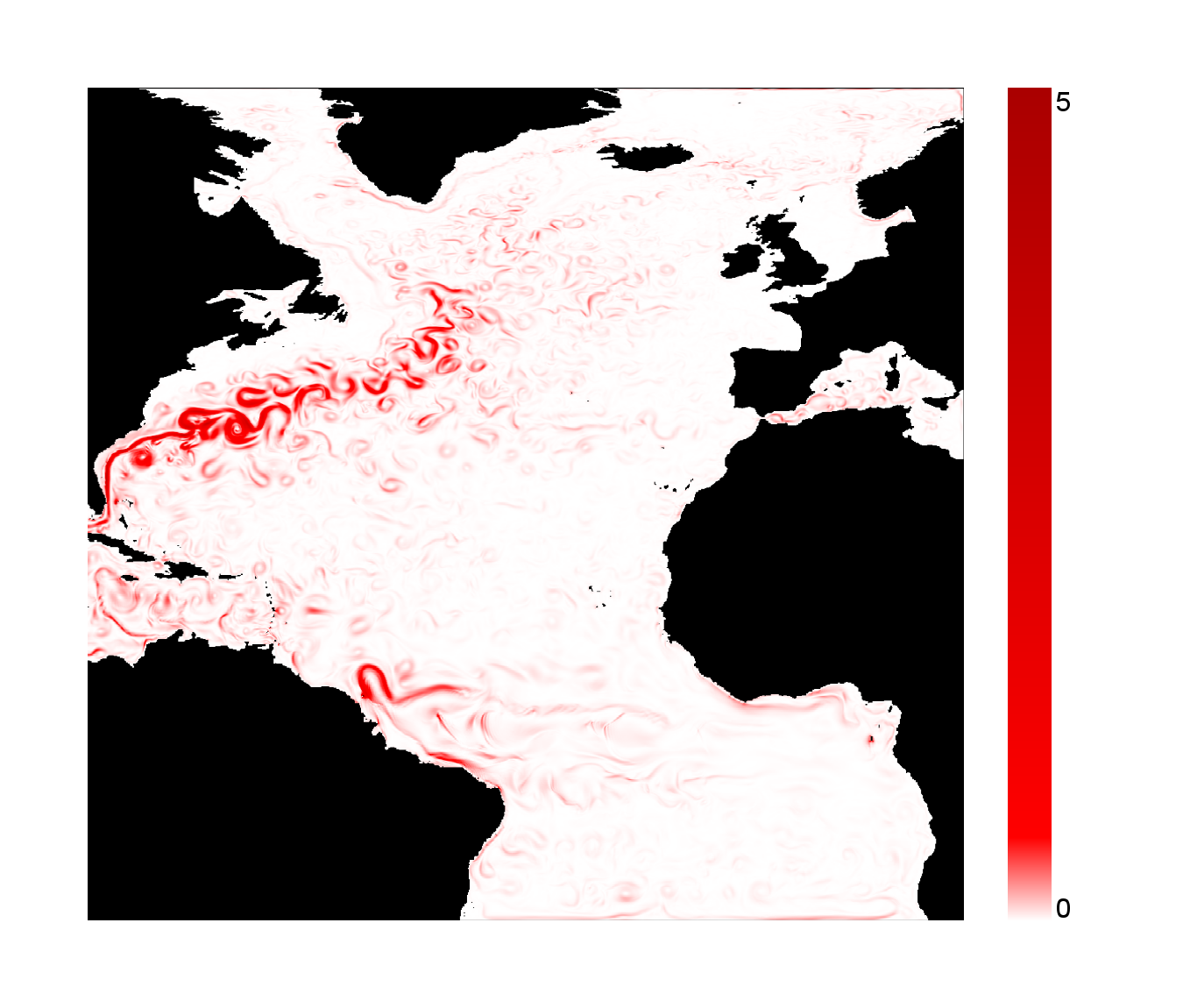}
\vspace{-.5cm}
	\caption{Our scale decomposition applied to model output (see Sec. \ref{sec:results}) at a single instant of time. Left column shows KE $|\OL{\bu}_\ell|^2/2$ (divided by density, in units of m$^2$/s$^2$), at scales larger than filtering scale $\ell$ (eq. \ref{def:filtering}). Right column shows KE, $|\bu'_\ell|^2/2$, in the complementary small-scales (eq. \ref{def:high-pass}). Rows show different filtering scales: (top) unfiltered with $\ell=0$ km, (middle) filtered  at scales $\ell=100$ km and (bottom)  $\ell=500$ km. Note the order of magnitude change in color scale to show energy below $\ell=100$ km (right middle panel). When visualizing in this manner, it is important to ensure the grid has sufficient resolution before taking the square of velocity to avoid aliasing effects.} 
	\label{fig:FilteringKnEn}
\end{figure}

The scale decomposition in (\ref{def:filtering}), (\ref{def:high-pass}) is
essentially a partitioning of scales in the system into large
($\gtrsim\ell$) and small ($\lesssim\ell$). Such a decomposition of 
the instantaneous flow in the North Atlantic into two sets of scales is 
shown in Figure \ref{fig:FilteringKnEn}, which makes plain two key 
advantages of the method: (i) an ability to vary the partitioning scale
$\ell$ to gain insight into the geographic location of different oceanic 
flow structures, and
(ii) an applicability to single snapshots, thereby allowing, for example, the generation of
movies of the flow at any set of length-scales.

While the simultaneous resolution of both spatial and scale 
information of a field $\bu(\bx)$ afforded by filtering is useful,  other 
decompositions such as with wavelet transforms can serve a similar
 purpose. (In fact, wavelets can be used within our 
approach with the proper choice of filtering kernel $G_\ell(\br)$.)
Other studies have used filtering for scale decomposition of oceanic data 
(e.g. \citep{ONeilletal12,Gaubeetal15}).
However, the true potential of the coarse-graining approach as an analysis 
framework derives mostly from utilizing the dynamical equations which 
describe the evolution of various scales. To do so, it is crucial to ensure that
the filtering operation (\ref{def:filtering}) commutes with spatial derivatives.
For example, it must satisfy $\grad\bdot\OL\bu_\ell = \OL{\grad\bdot\bu}_\ell$, which guarantees
that the filtered flow is incompressible if the original flow satisfies this property
as is the case for the flow in Figure \ref{fig:FilteringKnEn}. A simple
low-pass filtering, for example by averaging values at adjacent grid-cells 
or block-averaging on the sphere, does not satisfy these conditions and cannot be used for 
analyzing the dynamics at different scales as we do here.

The decomposition we use here preserves the fundamental physical 
properties of the flow, such as its incompressibility, its geostrophic character, 
and the vorticity present at various scales. This allows for the systematic and rigorous
derivation of equations governing any set of scales. For example, since our filtering 
commutes with spatial derivatives it mathematically guarantees that if one 
(i) filters the sea-surface height (SSH) field first, then computes the velocity or 
(ii) computes the velocity first, then filters it, the resultant coarse-grained velocity 
would be identical.

\subsection{Coarse-grained dynamics and scale-coupling}
Coarse-grained dynamical equations can be derived to describe the evolution of $\OL\bu_\ell(\bx)$ at every point $\bx$ in space and at any instant of time. For example, if $\bu(\bx)$ is governed by the rotating Boussinesq equations, then $\OL\bu_\ell(\bx)$ is governed by
\begin{equation} 
	\frac{\partial}{\partial t} \OL\bu_\ell + \OL\bu_\ell \bdot\grad \OL\bu_\ell = -\frac{1}{\rho_0}\grad\OL{P}_\ell -\bF\times \OL{\bu}_\ell -\grad\bdot\OL\tau_\ell(\bu,\bu) +\nu \nabla^2 \OL\bu_\ell +\frac{\OL{\rho}_\ell}{\rho_0} {\bf \,g}+ \OL{\bf F}_\ell^{\mbox{\scriptsize{forcing}}}.
\lb{largeMomentum}
\end{equation} 
Here, $P$ is pressure, $\bF$ is the Coriolis frequency, $\nu$ is viscosity, $\rho_0$ is the reference density, and $\bf F^{\mbox{\scriptsize{forcing}}}$ is forcing such as from winds or tides.
Eq. (\ref{largeMomentum}) is identical to the original unfiltered equation but with an additional contribution from the \emph{sub-filter stress} (oftentimes called ``subgrid stress'' in the LES literature), 
\be
\OL\tau_\ell(\bu,\bu) = \OL{\bu\bu}_\ell - \OL\bu_\ell\OL\bu_\ell~,
\lb{def:subgridstress}\ee 
a tensor representing the forces exerted by scales smaller than $\ell$ on the larger scale flow\footnote{The (traceless part of the) term $\OL\tau_\ell(\bu,\bu)$
is often thought of as a linear diffusive process and modeled as $-2\nu_{\mbox{{\scriptsize turb}}} \OL{{\bf S}}_\ell$, where $\bf S$ is the symmetric flow strain tensor. It is important, however, to remember that this is only a model which is often deficient and may sometimes fail altogether.} at every location $\bx$. In a Navier-Stokes flow, the sub-filter term $\OL\tau_\ell(\bu,\bu)$ contains all information needed to quantify the momentum coupling between the two sets of scales, $>\ell$ and $<\ell$. If we have complete knowledge of the dynamics in a simulated or real-life flow, {\it i.e.} knowing the velocity at every grid-point, the sub-filter stress can be calculated exactly at every point $\bx$ in the domain and at any instant in time t. Furthermore, since eq. (\ref{largeMomentum}) describes scales $>\ell$, for arbitrary $\ell$ (see Fig. \ref{fig:Filtering}), we can analyze the spatially-resolved nonlinear coupling as a function of scale $\ell$.

From the large-scale momentum equation (\ref{largeMomentum}), one can derive a KE budget for scales $>\ell$,
\begin{equation} \frac{\partial}{\partial t} \rho_0 \frac{|\OL\bu_\ell|^2}{2} + \grad\bdot\bJ_\ell^{\mbox{\scriptsize {transport}}} = -\Pi_\ell - \rho_0\,\nu\left|\grad\OL{\bu}_\ell\right|^2 +\OL{\rho}_\ell {\bf \,g}\bdot\OL\bu_\ell+ \rho_0\OL{\bf F}_\ell^{\mbox{\scriptsize{forcing}}}\bdot\OL\bu_\ell~~.
\lb{largeKE}\end{equation}
See, for example, \citet{Germano92} for details. Note that what we dub \emph{large-scale KE} is the KE in the large-scale flow, based on $\OL{\bu}_\ell$, rather than the filtered KE itself, $\rho_0 \OL{|\bu|^2}/2$, which \emph{does not} cascade across scales \citep{Germano92,Eyink05}. Here, 
\be\bJ_\ell^{\mbox{\scriptsize {transport}}}(\bx) = \rho_0 \frac{\left|\OL\bu_\ell\right|^2}{2}\OL\bu_\ell + \OL{P}_\ell\OL\bu_\ell - \rho_0 \nu \grad\frac{\left|\OL\bu_\ell\right|^2}{2} +\rho_0\,\OL\bu_\ell\bdot\OL\tau_\ell(\bu,\bu)
\nonumber\ee
represents the spatial transport of large-scale KE: the first term is advection by $\OL\bu_\ell$, the second is transport due to pressure, the third is diffusion due to molecular viscosity, and the last term accounts for the role of motion at scales $<\ell$ in transporting KE. The second term on the right hand side (RHS) of eq. (\ref{largeKE}) is direct destruction of large-scale KE by molecular viscosity and can be shown mathematically to be negligible at scales $\ell\gg\ell_d$ (e.g. \citep{Eyink08,Aluie13}). The third term is conversion from gravitation potential into kinetic energy, the analysis of which yields insight into baroclinic conversion that is believed to drive mesoscale eddies as we shall show in a follow-up work \citep{Sadeketal17}. The last term accounts for the direct kinetic energy injection due to forces such as wind or tides. The first term $\Pi_\ell$ is the energy scale-transfer or ``cascade'' 
term\footnote{The term `cascade' generally implies a spectrally-local transfer and, therefore, is a stronger statement than just `transfer,' although it is common in physical oceanography to use the two terms synonymously. However, in this manuscript, we henceforth avoid using the term `cascade' when unwarranted since we are not making any statement about the scale-locality of the transfer, which can be diagnosed through a more refined analysis similar to what was done in \citet{EyinkAluie09,AluieEyink09}.} and measures energy transferred from scales $>\ell$ to smaller scale due to nonlinear interactions. This is defined as
\begin{equation}
	 \Pi_\ell(\bx) = -\rho_0\OL\bS_\ell \bdots \OL\tau_\ell(\bu,\bu),
\lb{SGSflux}
\end{equation}
which is the large-scale strain tensor, $\OL\bS_\ell = (\grad\OL\bu_\ell + \grad\OL\bu_\ell^{\,\mbox{{\scriptsize T}}})/2$, acting against sub-filter scale stress, $\OL\tau_\ell(\bu,\bu)$. Here, the colon `$\bdots$' is a tensor inner product which yields a scalar. In a Navier-Stokes flow, $\Pi_\ell(\bx)$ contains all information needed to quantify the exchange of energy between the two sets of scales, $>\ell$ and $<\ell$. Since we have complete knowledge of the dynamics at all scales resolved in a simulation, $\Pi_\ell(\bx)$ can be calculated exactly at every point $\bx$ in the domain and at any instant in time $t$. This is demonstrated in Figure \ref{fig:SGSvsFrischGalilean}. It is not possible from simulation, satellite, or field data to capture all scales present in the real ocean. Therefore, computing $\Pi_\ell$ is only measuring the dynamical coupling between scales present in the data. It is possible to refine the analysis above by deriving an energy budget within a band of scales as was shown in \citet{EyinkAluie09}, however, the current analysis will suffice for the purpose of this paper.

While spatial maps of $\Pi_\ell(\bx)$ unravel a wealth of information about the scale dynamics, it is sometimes more insightful  to reduce such information by averaging over regions and plotting $\langle\Pi_\ell\rangle$ as a function of the remaining variable, $\ell$. Figure \ref{fig:SGSFlux_LVL01} shows an example of $\langle\Pi_\ell\rangle$ (plotted as a function of $1/\ell$ to make comparison to previous studies easier) which indicates the amount and sense of energy being transferred across different scales.

\subsection{Proper measure of the cascade}

In many instances, standard tools that were developed and used in the study of turbulence are only strictly valid to analyze homogeneous isotropic incompressible flows. Consequently, calculations of the energy transfer rates in the ocean that use these tools may give ambiguous results for inhomogeneous flows, as we show in Figure \ref{fig:SGSvsFrischGalilean}.  The problem arises because there are several possible definitions for the cascade term, $\Pi_\ell(\bx)$, in eq. (\ref{largeKE}), as we now elaborate.

Definition (\ref{SGSflux}) for the scale-transfer of energy in budget (\ref{largeKE}), which we shall call the \emph{sub-filter scale flux} or \emph{SFS flux}\footnote{The term ``flux'' in this context denotes a flux of energy across scales, which has units of power per unit volume. It should not be confused with a spatial flux, such as $\bJ_\ell^{\mbox{\scriptsize {transport}}}$ in eq.(\ref{largeKE}), which has units of power per unit area. Our terminology is borrowed from the turbulence and LES literature.}, is widely used in the LES literature (where $\Pi_\ell(\bx)$ is often called the ``subgrid scale flux'' or ``SGS flux'') but it is not unique. Another widely used definition is that applied to the ocean by the aforementioned studies, and that was largely developed in the context of homogeneous turbulence (HT) \citep{Frisch95} is
\begin{equation}
\Pi_\ell^{\mbox{{\scriptsize HT}}}(\bx)= \rho_0\,\bu\bdot(\grad\bu'_\ell)\bdot\OL{\bu}_\ell,
\lb{eq:FrischDef}
\end{equation}
where `$\bdot$' is a dot product between a tensor and a vector, which yields a vector. Yet a third possible definition, $\Pi_\ell^{\mbox{{\scriptsize uns}}}(\bx)=\rho_0\,[\grad\bdot(\OL{\bu\bu}_\ell)]\bdot\OL{\bu}_\ell$, which we shall refer to as the ``unsubtracted flux,'' was used by \citet{Lindborg06,Brethouweretal07} and \citet{MolemakerMcWilliams10} in idealized geophysically relevant flows.
The difference between any two of these definitions is a  divergence term,
$\grad\bdot(\dots)$, which amounts to a reinterpretation of which terms in budget (\ref{largeKE}) represent transfer of energy across scales and which terms redistribute (or transport) energy in space, 
$\grad\bdot\bJ_\ell^{\mbox{\scriptsize {transport}}}$. 
There is an infinite number of ways to reorganize terms in budget (\ref{largeKE}) and, thus, and infinite number of possible definitions for the transfer of kinetic energy between scales. This freedom in defining $\Pi_\ell(\bx)$ can be thought of as a \emph{gauge freedom}. 

In a homogeneous flow, spatial averages of all these definitions are equal because their difference is a divergence that is zero, $\langle\grad\bdot(\dots)\rangle=0$. On the other hand, if one considers inhomogeneous flows, such as in the ocean or atmosphere, or if one wishes to analyze the cascade geographically without spatial averaging, then such definitions can differ qualitatively as well as quantitatively. We will now argue that the SFS flux definition (\ref{SGSflux}) is the proper measure of the cascading energy because it satisfies an important physical criteria: {\itshape Galilean invariance}. Using such a criterion to choose the definition of the SFS flux may be thought of as \emph{gauge fixing}.

\subsubsection{Galilean invariance}

Galilean invariance is the requirement that a determination of the amount of energy cascading at any given point $\bx$ should not depend on the velocity of the observer. In other words, a measurement from a ship sailing in the Gulf Stream, and another from a station on land should register the same amount of energy being exchanged between scales.
\cite{Kraichnan64,Speziale85,Germano92,Eyink05} all emphasized the importance of Galilean 
invariance in the context of turbulence and, more recently, \citet{EyinkAluie09,AluieEyink09} showed that Galilean invariance was necessary for the so-called ``scale-locality'' of the cascade. There are non-Galilean-invariant terms in our budgets (\ref{largeKE}) but, as is physically natural, they are all associated with spatial transport $\bJ_\ell^{\mbox{\scriptsize {transport}}}$ of energy.

\begin{figure}
\centering
	\includegraphics[width=.75\textwidth,height=0.2\textheight]{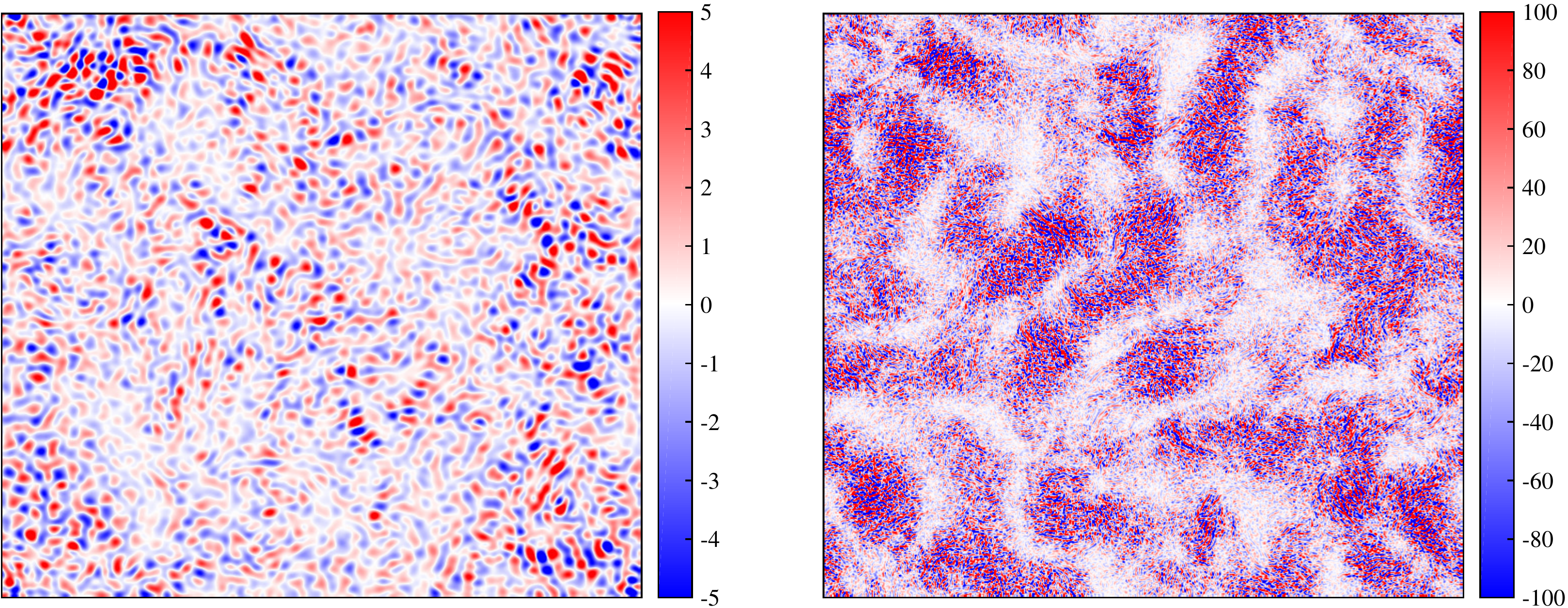}
	\includegraphics[width=.75\textwidth,height=0.2\textheight]{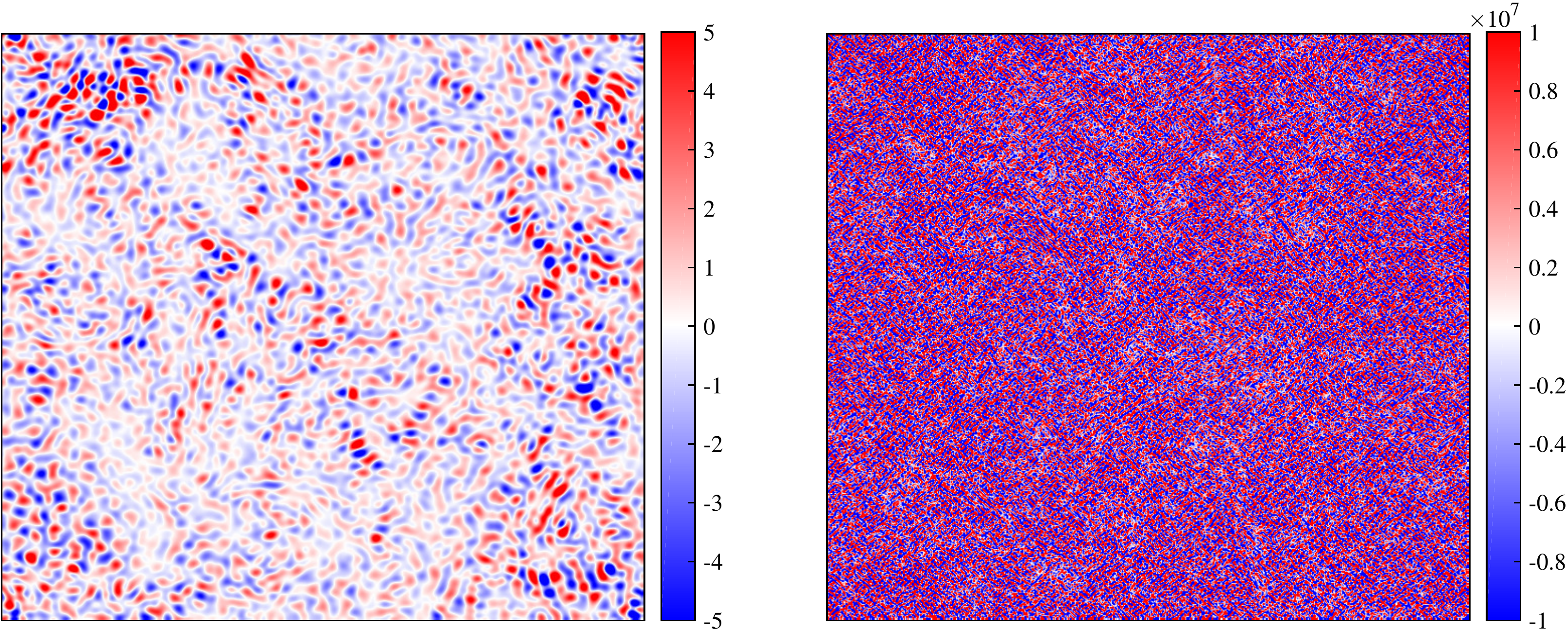}
     \caption{The energy transfer across scale $\ell=L/30$ ($L$ is the domain size), 
      at a single time-instant, at every $\bx$. Data is from a 3D triply periodic simulation 
      of homogeneous isotropic turbulence forced at large-scales.
      Red (blue) is energy transferred from scales
      larger (smaller) than $\ell$ to scales 
     smaller (larger) than $\ell$. Top-left panel uses the SFS cascade measure, which is Galilean
     invariant, whereas the top-right panel uses the HT definition \citep{Frisch95}, which is not, yielding
     an unphysical imprint of the large scales. 
     Bottom two panels measure the energy
     transfer using the respective definitions after embedding the fluctuations in a
     uniform mean flow, underscoring the dependence of  $\Pi_\ell^{\mbox{{\scriptsize HT}}}$ on 
     the observer's inertial frame of reference.
}
	\label{fig:SGSvsFrischGalilean}
\end{figure}

 Definition, $\Pi_\ell^{\mbox{{\scriptsize HT}}}(\bx)= \rho_0\,\bu\bdot(\grad\bu'_\ell)\bdot\OL{\bu}_\ell$, does not satisfy Galilean invariance. An observer moving at a constant velocity $-\bU_0$ relative to the system will measure a flux at point $\bx$, 
\be
 \Pi_\ell^{\mbox{{\scriptsize HT}}}(\bx)= \rho_0\,\left[\bu\bdot(\grad\bu'_\ell)\bdot\OL{\bu}_\ell 
+ \bu\bdot(\grad\bu'_\ell)\bdot \bU_0
+ \bU_0\bdot(\grad\bu'_\ell)\bdot \OL{\bu}_\ell
+ \bU_0\bdot(\grad\bu'_\ell)\bdot \bU_0\right],
\nonumber\ee
such that the amount of energy cascading at an arbitrary location in the flow will be dependent on the frame of reference. For large sweeping speeds, $|\bU_0|$, the measured cascade becomes proportional to $|\bU_0|^2$ as demonstrated in Figure \ref{fig:SGSvsFrischGalilean}. 

On the other hand, both the sub-filter stress in eq. (\ref{def:subgridstress}) and the SFS flux in eq. (\ref{SGSflux}), satisfy Galilean invariance. This property can be directly verified by the reader with elementary algebra, and is demonstrated in Figure \ref{fig:SGSvsFrischGalilean}.

The two cascade measures, visualized in Figure \ref{fig:SGSvsFrischGalilean}, show very different qualitative and quantitative behavior. $\Pi_\ell^{\mbox{{\scriptsize HT}}}$ has a strong dependence on the most energetic structures in the flow, with a conspicuous imprint of the strongest eddies forming the peak of the energy spectrum. 
To underscore Galilean invariance, we boost the velocity by a constant, $U_0\,\hat{\bx} = 1000\, \hat{\bx}$ m/s and recalculated the fluxes using both definitions. As expected, the SFS flux, $\Pi_\ell$, does not change. 
On the other hand, $\Pi_\ell^{\mbox{{\scriptsize HT}}}$ exhibits an unphysical dependence on the reference frame and is proportional to $O(U_0^2)$. 

The idea we are emphasizing is that any definition of a flux, which measures the amount of energy cascading across a scale $\ell$, should be Galilean invariant. Otherwise, the amount of energy cascading
at a point $\bx$ in the flow would depend on the inertial frame of reference of the system, which is unphysical.  Disentangling the cascade across scales from spatial transport is especially pertinent when trying to determine the sense of a cascade visually by looking at the evolution of structures. Consider a simple 3D pure Navier-Stokes turbulent flow in a laboratory tank. If one injects a localized blob of tracer somewhere in the flow, the blob will diffuse and expand. At face value, an observer might be tempted to conclude that the flow (and the tracer) is undergoing an inverse cascade because the size of the blob is growing. However, it is well-known that both energy and tracer variance in a 3D Navier-Stokes flow undergo a cascade to \emph{smaller} scales (e.g. \citep{Pope00}). The observed expansion is in fact due to spatial transport by turbulence, which is sometimes referred to as turbulent diffusion. If such spatial transport is subtracted by using a reference frame \emph{co-moving} with the \emph{local} large-scale flow,  then the observer would notice that the tracer, which started as a continuous blob, develops fine filaments and becomes fractal (down to the viscous scales) indicating a downscale cascade. This is precisely what the SFS flux definition (\ref{SGSflux}) measures. While Figure \ref{fig:SGSvsFrischGalilean} relies on a boosting velocity for the purpose of illustration, we will see below that Galilean invariance proves to be of significance in a number of regions within the North Atlantic Ocean.

\section{Results} \lb{sec:results}

Since  we can probe the dynamics simultaneously in scale,
$\ell$, and in space, $\bx$, we show two types of results below. The
first type explores the energy transfer across various scales $\ell$,
spatially averaging $\langle \Pi_\ell \rangle$ over a region of interest.
The region may be very small in geographic extent or very
large, encompassing the entire domain. The second type of results we
present keeps scale $\ell$, across which energy is being transferred,
fixed while fully resolving the scale-transfer in space. This yields spatial
maps of $\Pi_\ell(\bx)$.

As we mentioned above, it is important to bear in mind that the scale
coupling unveiled by our analysis is an exact description of the
fully nonlinear dynamics \emph{in the simulation}, which may be
different from that of the real ocean. This is a limitation shared by
any qualitative
or quantitative analysis done on data, be it from simulations or
observations.

We will now present an analysis of the scale-transfer in the North Atlantic
using OGCM data. Our method can also be applied to observational data,
including satellite altimetry and ARGO floats. But here we take
advantage of the uniform and complete coverage afforded through the
use of simulation output in order to establish the effectiveness of
the method. We hope that it will then subsequently be applied to 
observational data.

\subsection{Description of simulation}
The simulation we analyze is the 14b case of \citet{Bryanetal07}. As
explained in that paper, this simulation was generated with the POP
free surface, hydrostatic primitive equation code
\citep{DukowiczSmith94} using z-coordinates, and a full-cell
representation of topography. A mercator grid with zonal grid spacing
of 0.1$^{\circ}$ and meridional spacing of 0.1$^{\circ}$ x
sin(latitude) covered the Atlantic basin from 20$^{\circ}$S to
73$^{\circ}$N, including the Gulf of Mexico and the western
Mediterranean. Towards the northern boundary of the domain, 
the first internal Rossby radius of deformation becomes poorly 
resolved, with only one grid cell spanning the Rossby radius 
at the highest latitudes of the North Atlantic basin (see Fig. 1 of \citet{Smithetal00}).
This important dynamical length scale tends to be adequately 
resolved, however, at the latitudes of our analysis regions 
(see our Fig. 4 below for the analysis regions).
A 40-level grid was used in the vertical with cell
thickness increasing from 10 m at the surface to 250 m in the deep
ocean. Biharmonic eddy viscosity, $\nu$, and diffusivity, $\kappa$,
was used, scaled with the cube of the local grid spacing, as \be \nu =
\nu_0\left(\frac{dx}{dx_0}\right)^3 \ee (and similarly for $\kappa$),
such that the grid-scale Reynolds number \be Re_{\mbox{\scriptsize{grid}}} =
\frac{U dx^3}{\nu_0} \ee is constant for a fixed velocity scale,
regardless of location on the grid. Here, $\nu_0=-1.35\times 10^{10}$
m$^4/$s and the corresponding diffusive coefficient was $\kappa_0 =
\nu_0/3$. The \citet{PacanowskiPhilander81} parametrization of
vertical mixing was used with background values of viscosity and
diffusivity of $10^{-4}$ and $10^{-5}$ m$^2/$s, respectively. A
quadratic bottom stress with a drag coefficient of
$1.225\times10^{-3}$ was applied.

The experiment was forced with a daily averaged wind stress, computed
from ECMWF TOGA surface analyses (derived from operational forecasts
and provided on a 1.125$^{\circ}$ Gaussian grid) for mid-1985 to early
2001. A repeating annual cycle of surface heat flux is prescribed
using the Newtonian cooling boundary condition of
\citet{Barnieretal95} with a penetrative solar radiation flux. A
restoring boundary condition is used for surface salinity, damping the
model solution toward the \citet{Levitus82} monthly climatology on a
timescale of one month. The north and south boundaries of the domain
are closed to flow, with temperature and salinity restored to the
annual mean Levitus climatology within 3$^{\circ}$ wide buffer
zones. Details of this and other points of model configuration appear
in the earlier paper of \citet{Smithetal00}.

The model output that we analyze was saved at ten day intervals over a
three year period of the simulation, from March 1998 through February
of 2001.

\subsection{Mapping energy scale-transfer}
\begin{figure}
\centering
	\includegraphics[width=1\textwidth,height=0.7\textheight]{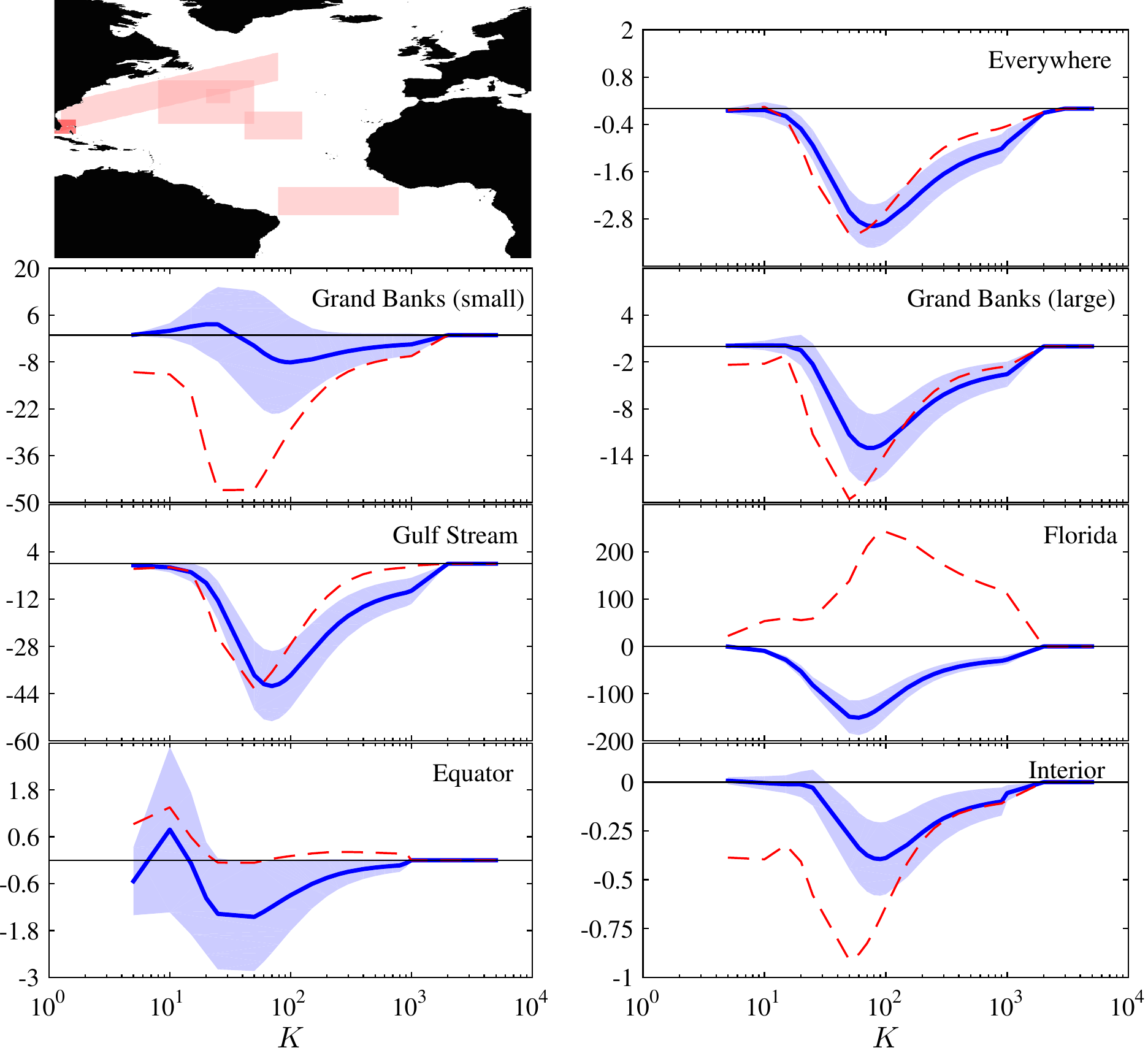}
	\caption{At the ocean surface: spatially averaged SFS flux (solid-blue line), $\langle\Pi_\ell\rangle$ (W/km$^2$/m), as a function of scale, $K=10^4/\ell$ km$^{-1}$. Here, $K$ is not a wavenumber, just a number proportional to $\ell^{-1}$. The uppermost left panel shows the various regions over which $\Pi_\ell(\bx)$ is averaged. The transparent blue shade depicts the temporal standard deviation in the SFS flux over a 3-year period (110 snapshots), while the solid blue line is the temporal average. This is compared to the homogeneous turbulence flux (dashed-red), $\langle\Pi_\ell^{\mbox{{\scriptsize HT}}}\rangle$, without artificial tapering or boosting. Significant qualitative (and not just quantitative) differences in Florida, the Equator, and the Grand Banks  where strong mean currents exist (Gulf stream and N. Equatorial current), sweeping through the box.}
	\label{fig:SGSFlux_LVL01}
\end{figure}

\begin{figure}
\centering
	\includegraphics[width=1\textwidth,height=0.7\textheight]{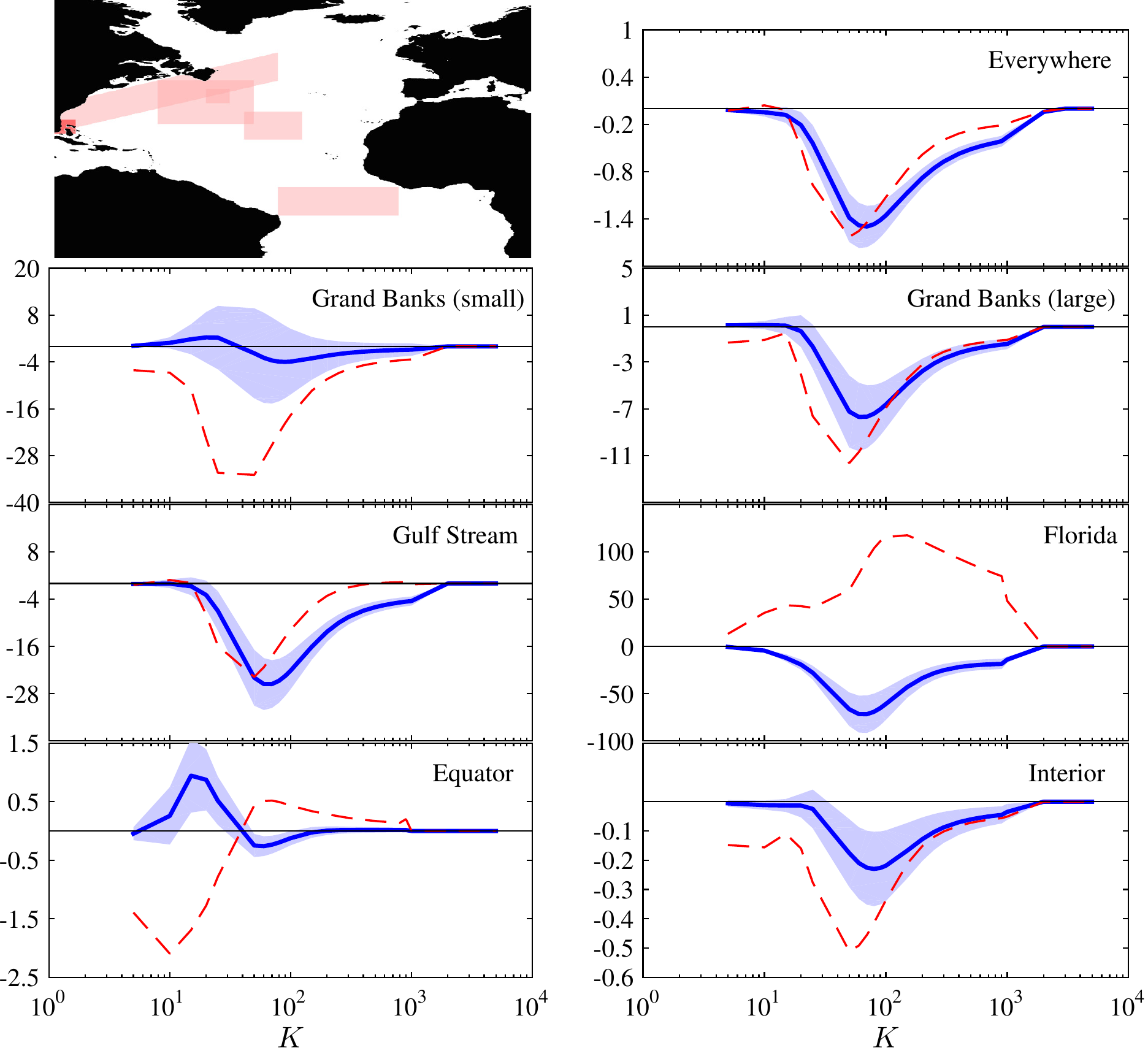}
	\caption{As in Figure \ref{fig:SGSFlux_LVL01} but at 100 m depth: $\langle\Pi_\ell\rangle$ (solid-blue line) and $\langle\Pi_\ell^{\mbox{{\scriptsize HT}}}\rangle$ (dashed-red). Units on the y-axis are in W/km$^2$/m. We again see significant differences between the two measures of energy transfer, especially in Florida, the Equator, and the Grand Banks  where 
strong mean currents sweep through the box.}
	\label{fig:SGSFlux_LVL10}
\end{figure}

\begin{figure}
\centering
	\includegraphics[width=1\textwidth,height=0.7\textheight]{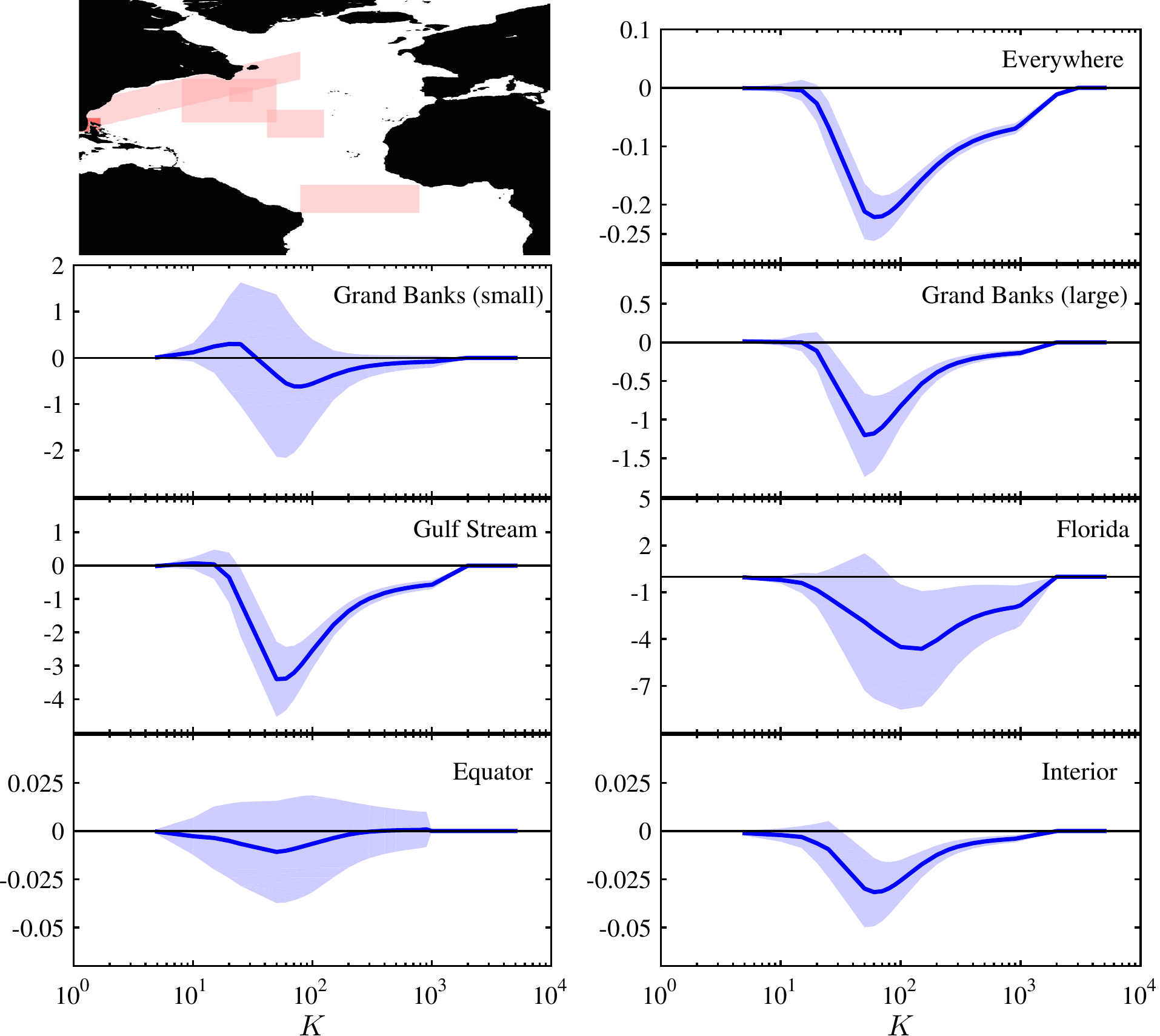}
	\caption{At 500 m depth: $\langle\Pi_\ell\rangle$ (solid-blue line)  as in the previous Figure \ref{fig:SGSFlux_LVL01}. Units on the y-axis are in W/km$^2$/m. This depth is within the thermocline, where stratification effects are, on average, strongest in the ocean. We notice that the energy transfer across scales is an order of magnitude smaller here than in the mixed layer, near the surface.}
	\label{fig:SGSFlux_LVL19}
\end{figure}

\begin{figure}
\centering
	\includegraphics[width=1\textwidth,height=0.7\textheight]{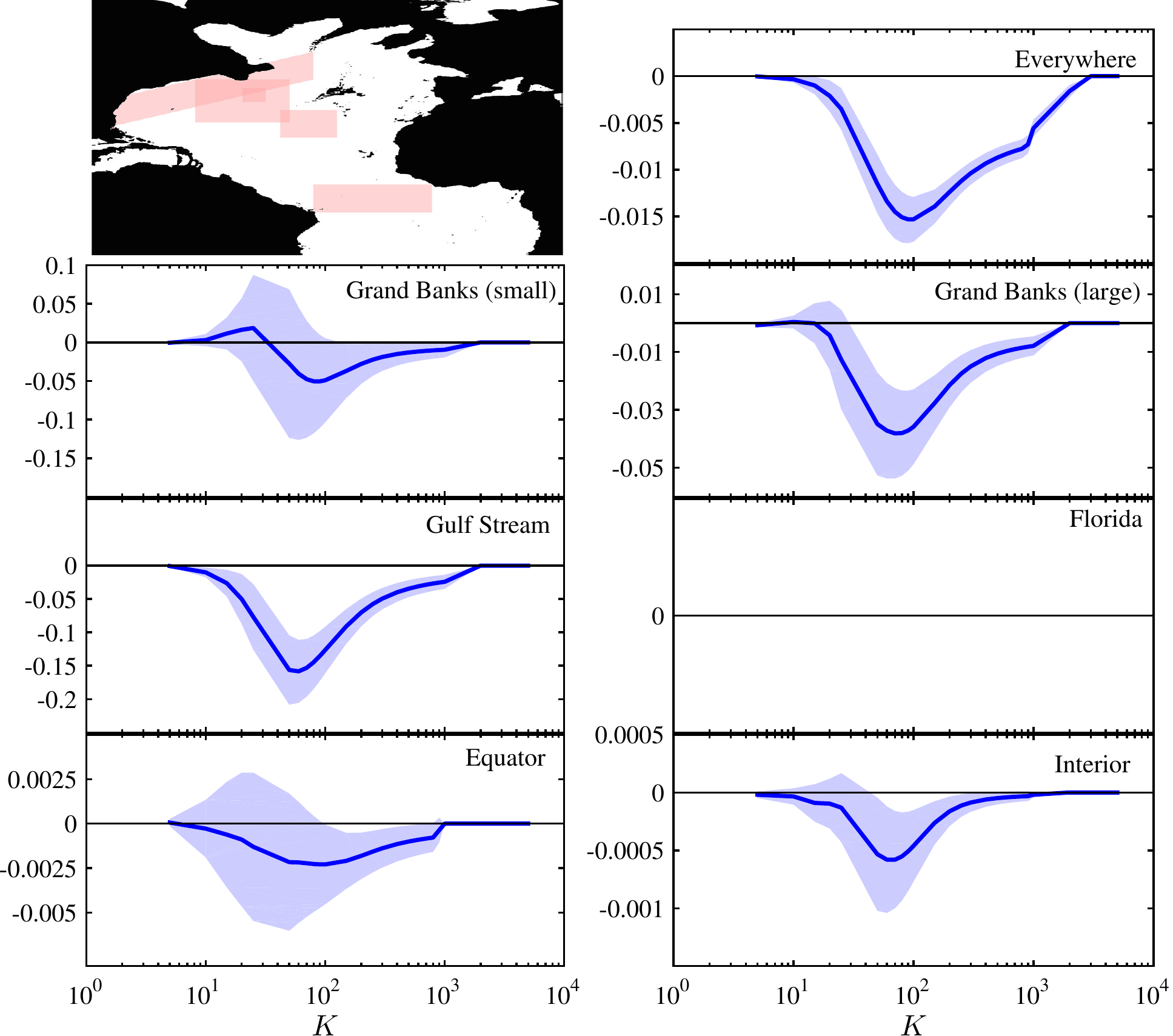}
	\caption{At 2,000 m depth: $\langle\Pi_\ell\rangle$ (solid-blue line)  as in the previous Figure \ref{fig:SGSFlux_LVL01}. Units on the y-axis are in W/km$^2$/m. This is within the deep ocean, where stratification effects are, on average, weakest in the ocean. We notice that the energy transfer across scales is two order of magnitude smaller than in the mixed layer, near the surface.}
	\label{fig:SGSFlux_LVL26}
\end{figure}

Figures \ref{fig:SGSFlux_LVL01}-\ref{fig:SGSFlux_LVL26} show the
scale-transfer at various locations and depths in the ocean as a function of
scale $\ell$. (The plot is a function of $1/\ell$ to make comparison to 
previous studies easier). The depths we have chosen are: (i) at the surface to
compare to previous results using altimetry data, (ii) at 100 m depth, 
slightly below the average depth of the bottom of the mixed layer, (iii) at 500 m
depth, within the thermocline, and (iv) at 2,000 m within the more
weakly stratified depths of the 
ocean. The regions we have chosen are: (i) the entire North Atlantic
domain of our data to characterize the scale-transfer at the
basin scale, (ii) a large region near the Grand Banks, in the Gulf Stream
Extension, which overlaps with the region studied in
\citet{Arbicetal13} to compare with that study, (iii) a smaller region
of the Grand Banks, to
test the role of region size on the dis/agreement between the SFS and
the HT cascade measures, (iv) a small region east of Florida, within
the Gulf Stream, where sweeping effects are very large, (v) a region
encompassing most of the Gulf Stream and its extension to measure the
scale-transfer in a western boundary current system, something which was absent from
previous studies and which can be carried out with our
coarse-graining approach, (vi) a small region in the Sargasso Sea, at
approximately the center of the North Atlantic's Subtropical Gyre, to measure the
scale-transfer in a relatively quiescent region within which
sweeping effects are negligible, (vii) an equatorial region,
between 5$^{\circ}$S -- 5$^{\circ}$N and 10$^{\circ}$W--35$^{\circ}$W,
which covers the North Equatorial Current and Counter Current in our
simulation, and where the geostrophic approximation fails.

Figures \ref{fig:SGSFlux_LVL01}-\ref{fig:SGSFlux_LVL10} show that
HT and SFS are qualitatively different in regions where sweeping
effects by mean oceanic current are important, especially in the
Florida region, but also in the Equator and the small Grand Banks
boxes. In the larger Grand Banks region, where the mean Gulf Stream is
relatively less dominant, SFS and HT are in qualitative
agreement. When averaging over the entire domain, over which mean sweeping
effects are exactly zero (zero flow in or out of the domain), the two
definitions yield very similar results.  From Figures
\ref{fig:SGSFlux_LVL01}-\ref{fig:SGSFlux_LVL10}, we also see that the
interior region in the Sargasso Sea is not as homogeneous as one may
think, where we
find that the HT and SFS definitions agree over scales smaller than 50 km, but
diverge over larger scales.

\begin{figure}
\centering
\vspace{-2cm}	
	\includegraphics[trim=1cm 1cm 1cm 1cm,clip, width=0.8\textwidth,height=0.4\textheight]{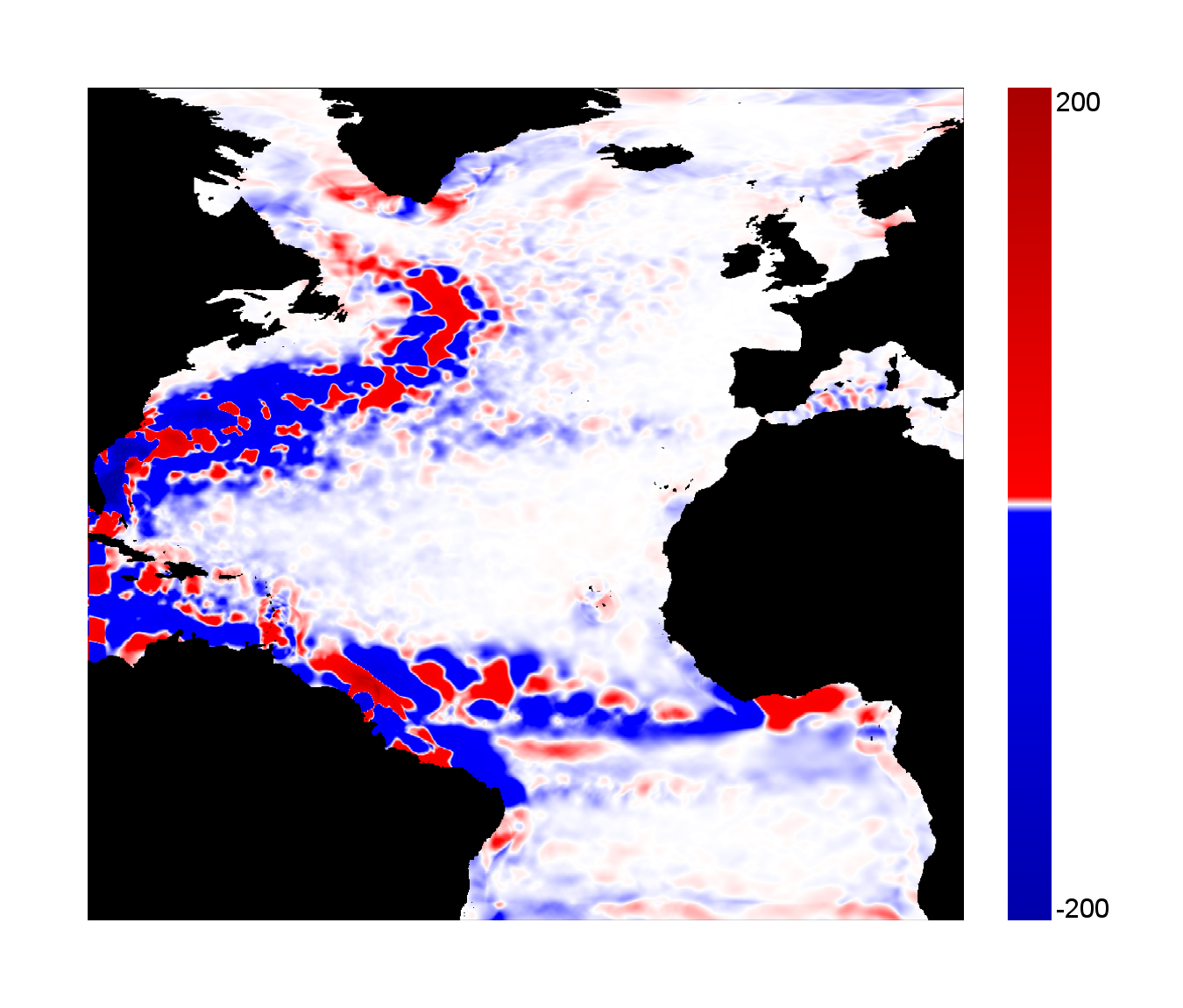}\vspace{-1cm}
	\includegraphics[trim=1cm 1cm 1cm 1cm,clip, width=0.8\textwidth,height=0.4\textheight]{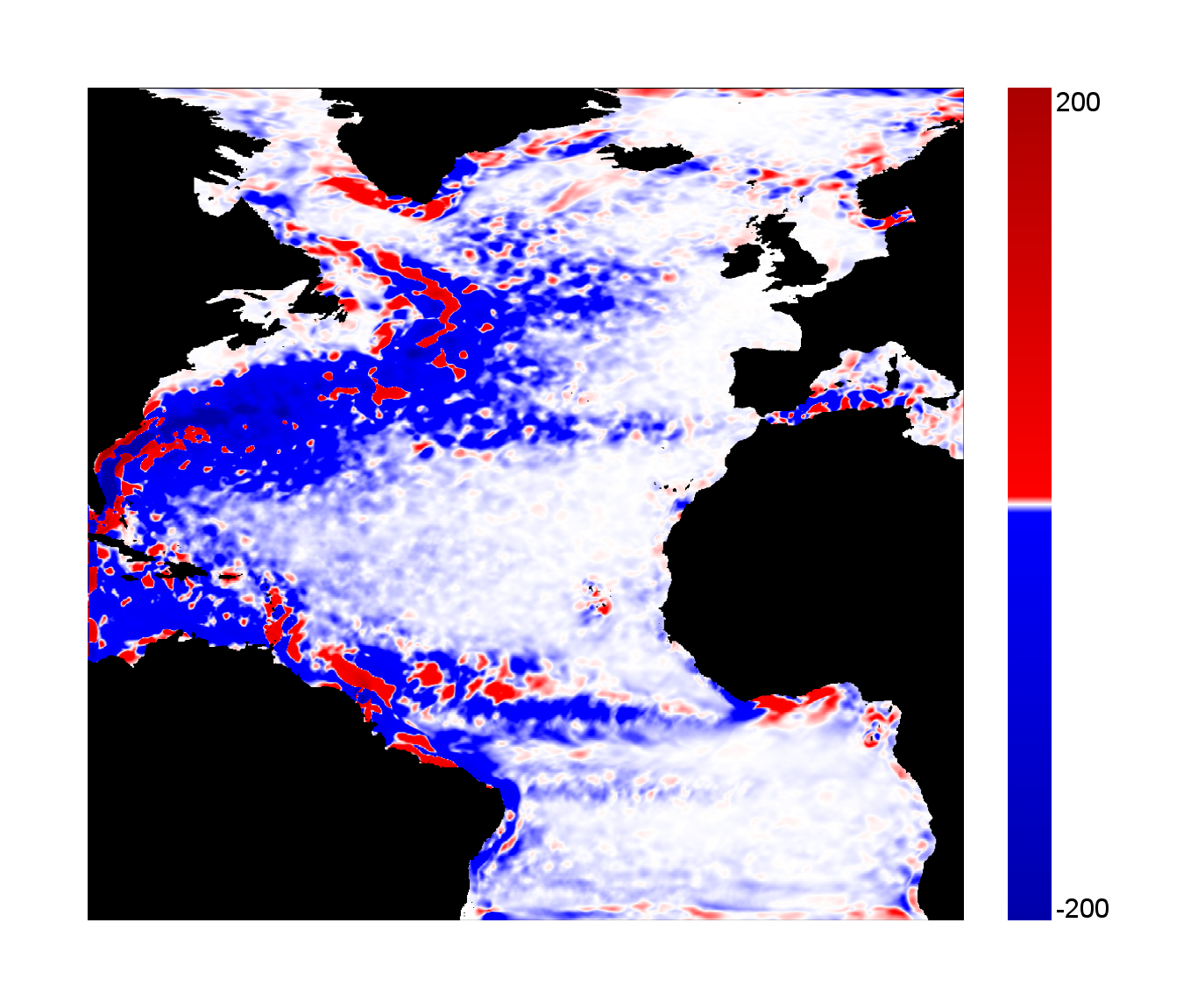}
\vspace{-0.5cm}	
	\caption{Geographic maps of the inter-scale energy transfer, $\Pi_\ell(\bx)$ (W/km$^2$/m), at the surface, time-averaged over 3 years (110 snapshots), where $\ell = 400$ km (top) and $\ell = 200$ km (bottom). The color map used, which has units of W/km$^2$/m, is not linear: most of the color shown has small values close to zero (white) and some blue/red regions exceed the maximum values on the color bar. We observe a downscale transfer in the current South of Florida, as the Gulf Stream turns northward, possibly indicative of eddy shedding or even just the small-scale associated with the sharp turn in the trajectory. We also observe a strong (dark blue) upscale transfer in the Gulf Stream core East of Florida and the Carolinas. This persists well beyond the separation point (Cape Hatteras), indicating that energy is transferred from mesoscale eddies into the Gulf Stream, accelerating and focusing the current. Flanking both sides of this (dark blue) core, we see downscale transfer (red) most probably associated with barotropic instabilities resulting from strong shear. Overall, an upscale transfer dominates in the Gulf Stream, in accord with QG. A similar pattern, though not as pronounced, exists in the N. Brazil Current. The (shallow) N. Equatorial Current, which in our simulation is around 5$^\circ$N, exhibits an upscale energy transfer.  
}
	\label{fig:SGSavgLVL01}
\end{figure}

\begin{figure}
\centering
\vspace{-1cm}	
	\includegraphics[trim=1cm 1cm 1cm 1cm,clip, width=0.8\textwidth,height=0.4\textheight]{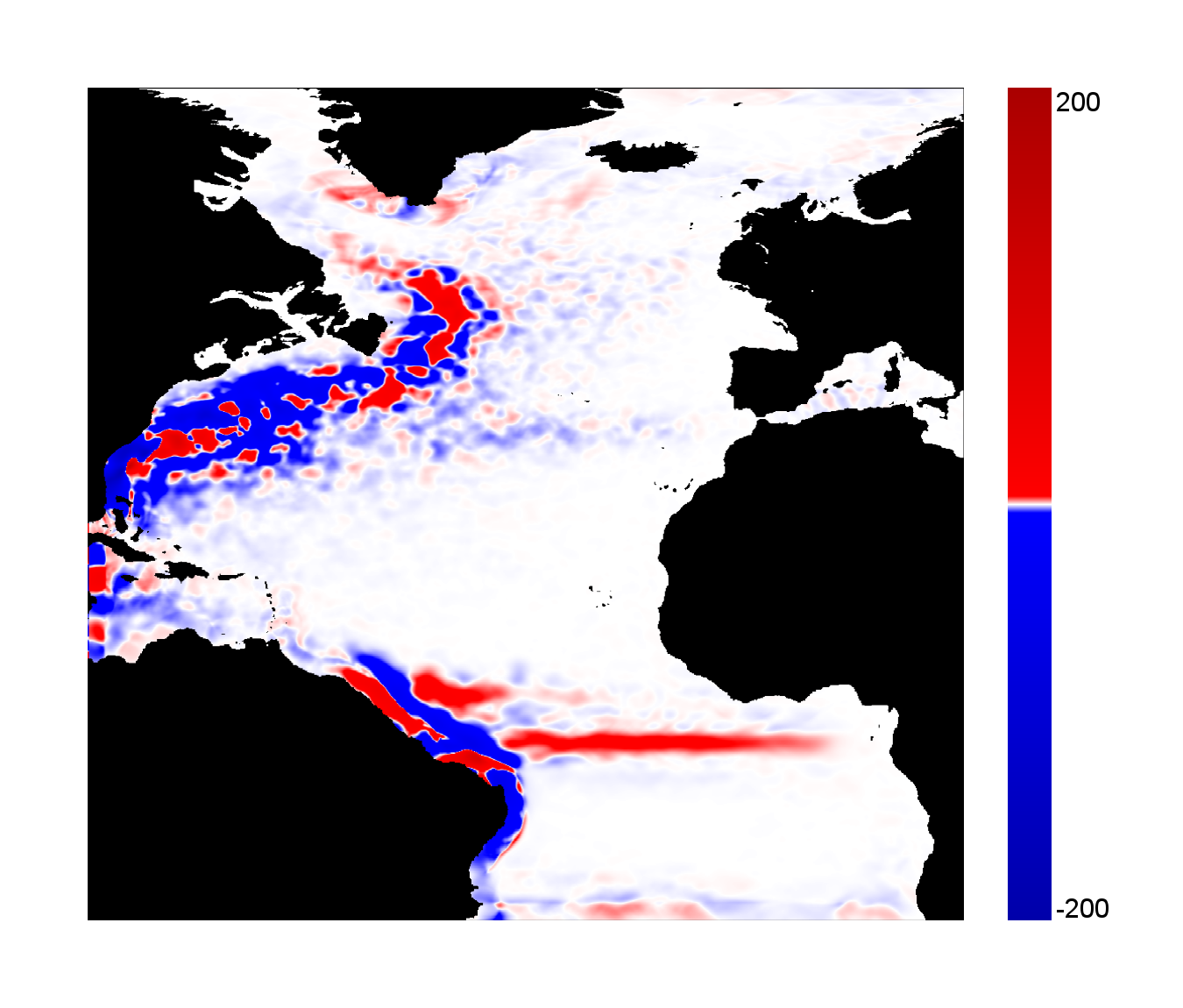}
\vspace{-0.5cm}	
	\includegraphics[trim=1cm 1cm 1cm 1cm,clip, width=0.8\textwidth,height=0.4\textheight]{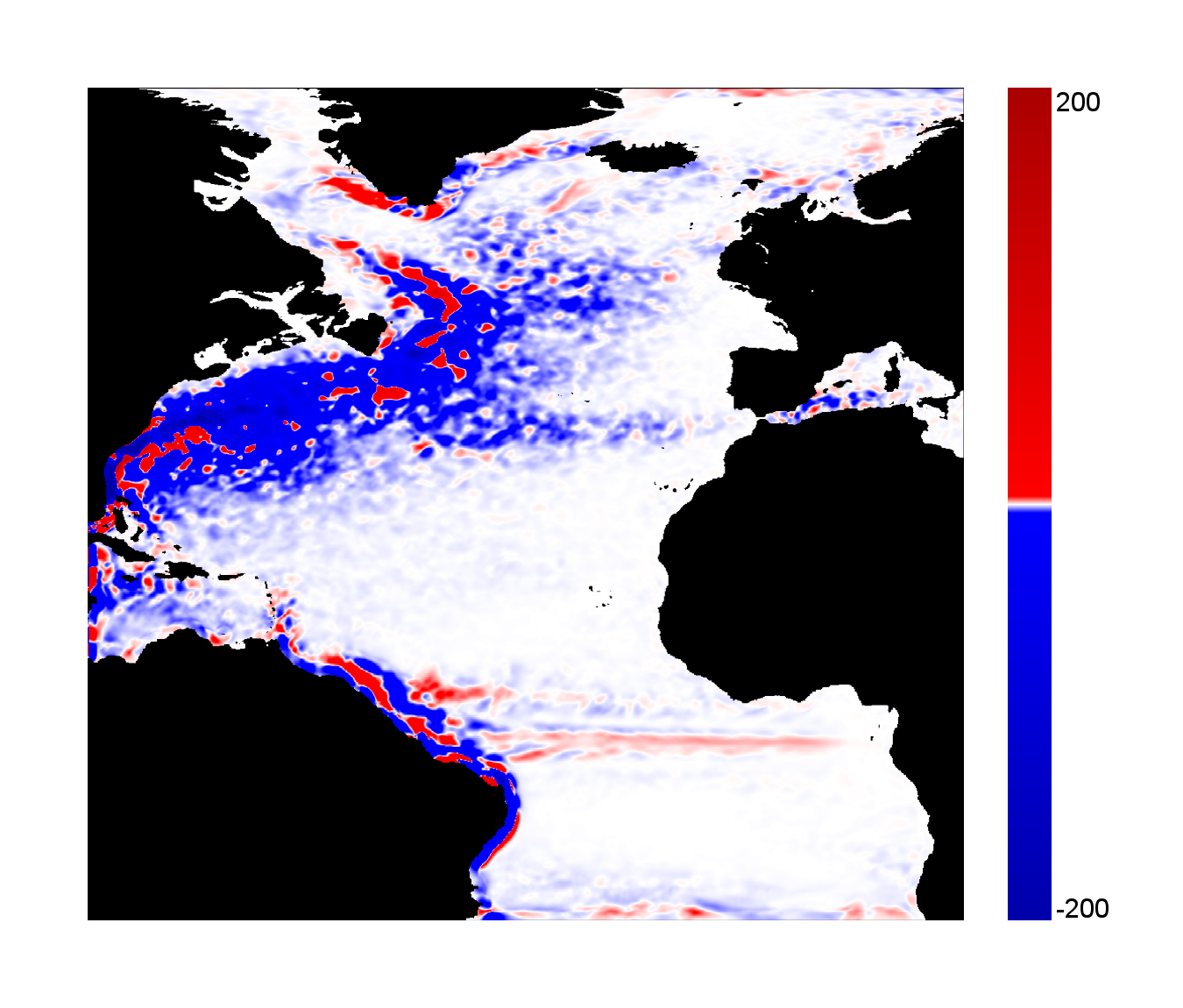}
	\caption{Maps of $\Pi_\ell(\bx)$, as in Figure \ref{fig:SGSavgLVL01}, but at 100 m depth, where $\ell = 400$ km (top) and $\ell = 200$ km (bottom). We notice in the Gulf Stream a pattern similar to that at the surface. In fact, almost the exact red/blue patch patterns that appear at the Gulf Stream surface appear at 100 m and 500 m depth (see next Fig. \ref{fig:SGSavgLVL19-26}), suggesting that $\Pi_\ell(\bx)$, as a scalar field, is depth-independent at high latitudes. On the other hand, we notice that there is a downscale transfer of energy in the Equatorial Counter Current, which in our simulation, is approximately at 0$^\circ$N and 100 m depth,  indicating an obvious departure from the QG model.
}
	\label{fig:SGSavgLVL10}
\end{figure}

\begin{figure}
\centering
	\includegraphics[trim=1cm 1cm 1cm 1cm,clip, width=0.8\textwidth,height=0.4\textheight]{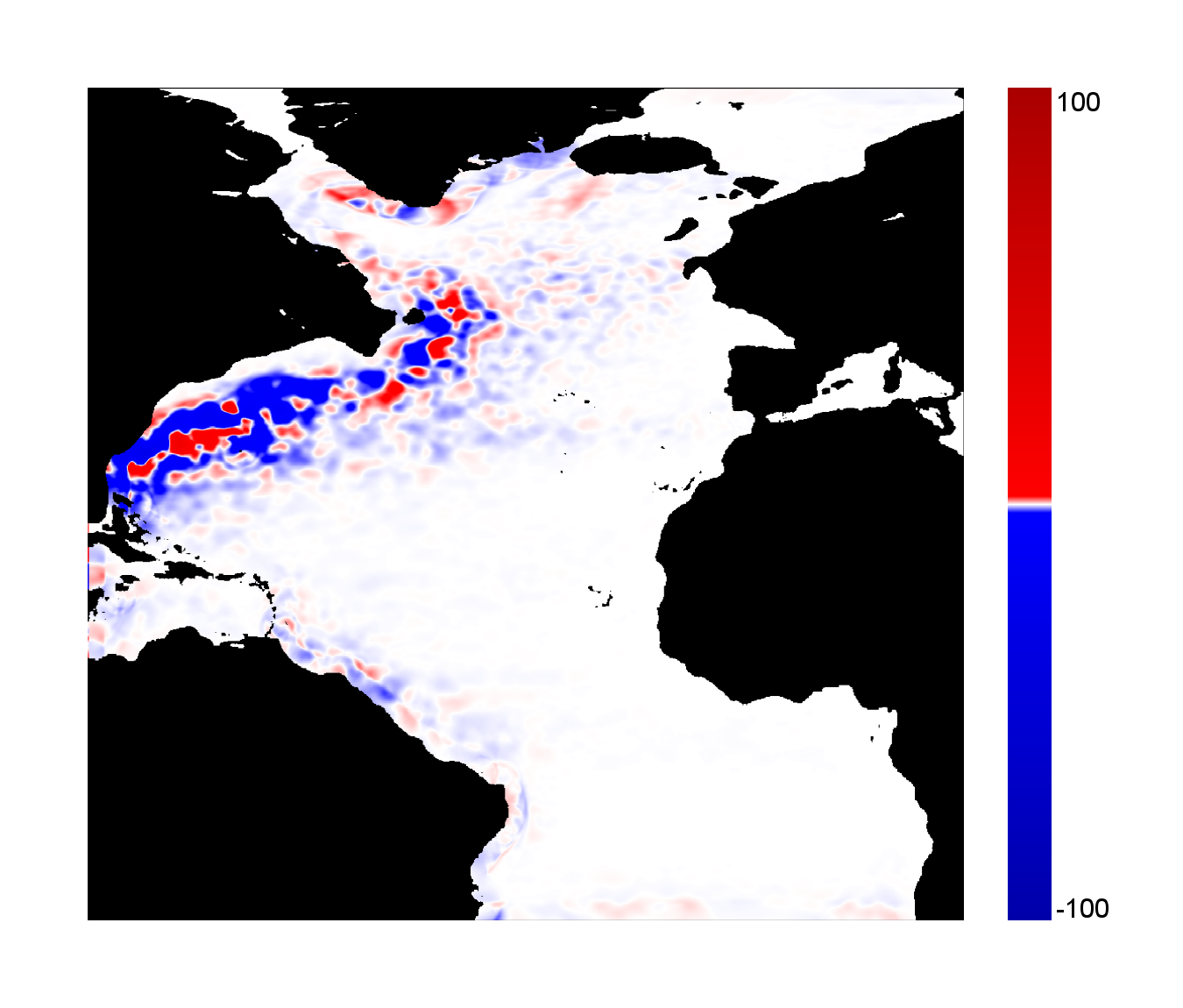}
	\includegraphics[trim=1cm 1cm 1cm 1cm,clip, width=0.8\textwidth,height=0.4\textheight]{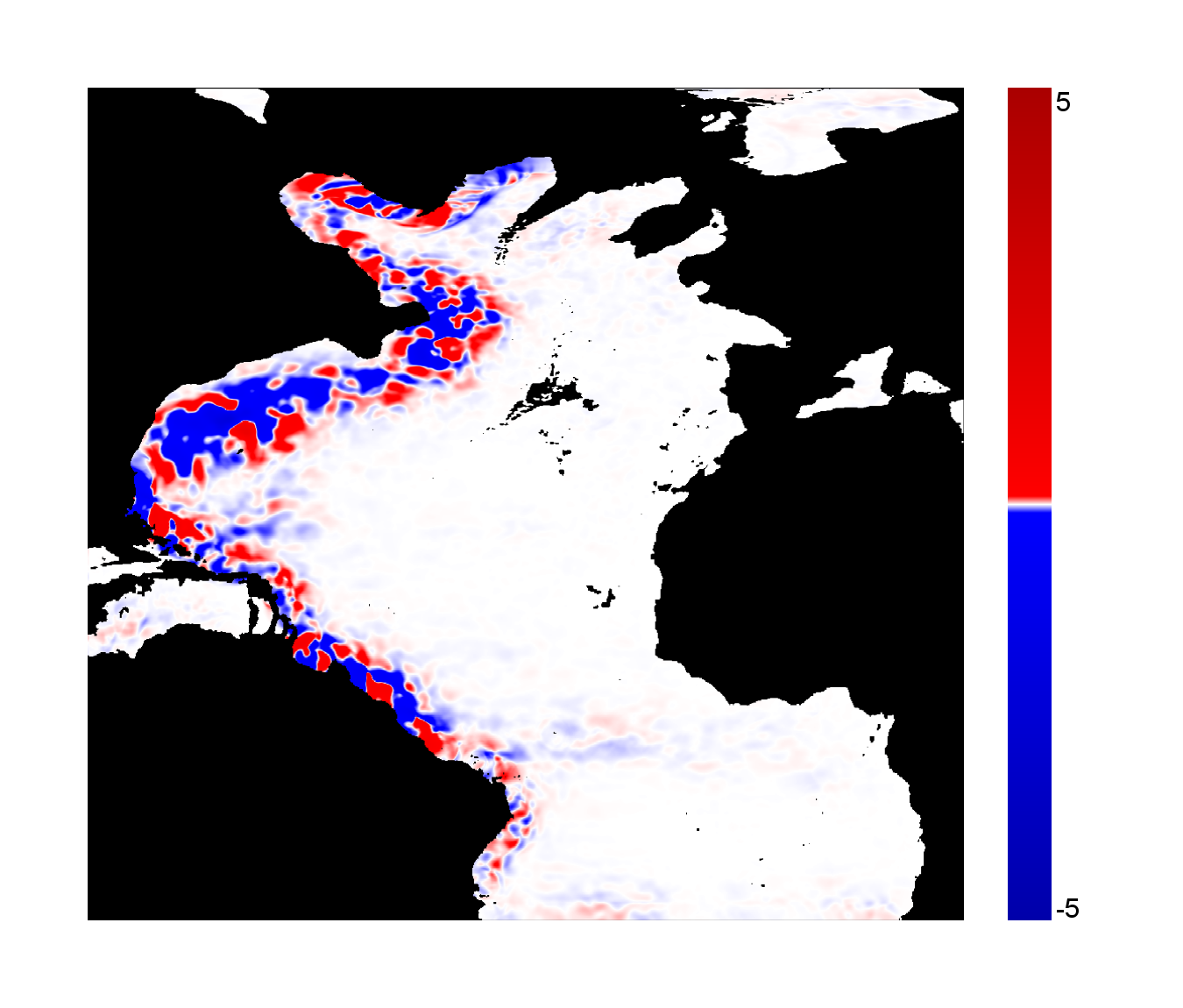}
\vspace{-0.5cm}	
	\caption{Map of $\Pi_\ell(\bx)$, as in Figure \ref{fig:SGSavgLVL01}, but at depths of 500 m (top) and $2,000$ m (bottom), where $\ell = 400$ km.  We notice that within the thermocline (top), scale-transfer is weak with the exception of the Gulf Stream which is known for its deep penetration. The red/blue patch patterns are similar to those appearing at the Gulf Stream surface and at 100 m. It is also clear from the color bar magnitudes of the bottom panel that there is relatively weak scale-transfer in the deep ocean.}
	\label{fig:SGSavgLVL19-26}
\end{figure}

Figures \ref{fig:SGSavgLVL01}-\ref{fig:SGSavgLVL19-26} show geographic 
maps of inter-scale energy transfer in the ocean. They are time-averaged over 
a 3-year period, which we have checked to be almost identical to 2-year averaged
maps, indicating that the features shown are persistent in time.
The maps reveal intense 
KE scale-transfer taking place in the Gulf Stream and
in the North Brazil Current. There is also weaker but
significant scale-transfer in most of the North Atlantic, not as
visible due to the color map. As one would expect, the scale-transfer
is significantly stronger in the uppermost layers compared to the
deeper ocean. 

The qualitative nature of scale-transfer differs significantly at various 
geographic locations and depths. For example, we observe an upscale 
transfer of energy in the surface equatorial region, where the 
equatorial flow is from east to west. On the other hand, 
Figs. \ref{fig:SGSavgLVL10} and \ref{fig:SGSFlux_LVL01} show that 
in the Equatorial Counter Current at 100 m depth, the transfer is almost 
entirely downscale. We note that QG theory is not expected to hold at 
the Equator.

Another prominent feature we observe in Figures \ref{fig:SGSavgLVL01} 
and \ref{fig:SGSavgLVL10}, especially across $\ell=200$ km at which 
the transfer peaks, is a strong (dark blue) upscale transfer in the Gulf 
Stream core east of Florida and the Carolinas. This persists well 
beyond the separation point (Cape Hatteras), indicating that energy is 
transferred from mesoscale eddies into the Gulf Stream, accelerating 
and focusing the current. Flanking both sides of this (dark blue) core, 
we see downscale transfer (red) most probably associated with 
barotropic instabilities resulting from strong shear. This supports 
recent eddy-mean flow interaction models which rely on decomposing 
the flow into mean and fluctuating parts (\citep{Klockeretal16}). 
Overall, an upscale transfer dominates in the Gulf Stream, in general 
accord with the traditional QG paradigm. A similar pattern, though not as pronounced, 
exists in the North Brazil Current. 

We also observe from Figures \ref{fig:SGSavgLVL01} and 
\ref{fig:SGSavgLVL10} a persistent red patch in the Gulf Stream as it 
passes through the Florida Strait just south of the peninsula. This 
indicates an expenditure of kinetic energy by the Gulf Stream as it 
traverses the Florida Straight and undergoes a sharp turn northward. 
This is not necessarily associated with a slowdown in the mean 
current speed since any loss may be offset by other forcing 
mechanisms, such as buoyancy or wind forcing. Another northward 
turn occurs at the Grand Banks, where the North Atlantic current 
carries subpolar gyre waters further poleward than anywhere else 
on Earth. Here, a strongly coherent red core of downscale energy 
transfer is flanked on both sides by upscale transfer, or inverse 
cascade.  An in-depth investigation of these issues is worth pursuing in
future work but would take us beyond the scope of this paper.

Figures \ref{fig:SGSavgLVL01}-\ref{fig:SGSavgLVL19-26} 
show that the SFS flux, which is a scalar field, seems to be 
mostly depth-independent at high latitudes. In other words, the pattern of red versus
blue in the Gulf Stream, indicating the sense of the energy scale-transfer,
appears to be nearly the same at the surface, at 100 m depth, 
and also at 500 m depth, which has practical utility in 
inferring transfer from surface altimetry data.

We also notice from Figures\ref{fig:SGSFlux_LVL01}-
\ref{fig:SGSFlux_LVL26} that the temporal fluctuations in the 
scale-transfer differ as a function of the averaging box and the nature of 
the flow within the box. For example, we find that in regions where a 
relatively strong coherent mean flow exists, such as in the 
equatorial counter current at 100 m depth or near Florida, the 
temporal variation is smaller than in regions which lack a strong 
coherent mean flow, such as at the equatorial surface, Sargasso 
Sea, and the small Grand Banks region where the instantaneous 
sweeping by the Gulf Stream Extension is strong but is not 
temporally coherent.

A general conclusion we can deduce from Figures
\ref{fig:SGSavgLVL01}-\ref{fig:SGSavgLVL19-26} is that an upscale energy transfer
does not take place everywhere in the ocean, even at the higher latitudes.
On the other hand, if we average over large enough regions (of order $10^3$ km in size or larger) in
the ocean, away from the equator, we find from Figures
\ref{fig:SGSFlux_LVL01}-\ref{fig:SGSFlux_LVL26} that the quasi-geostrophic paradigm
of an upscale transfer is a qualitatively correct \emph{mean description} of the 
energy scale-transfer in a basin-averaged sense.

It is well-documented in the homogeneous isotropic turbulence
literature (\citep{Chenetal03,Boffetta07,Riveraetal14}) that the
spatial distribution of the SFS flux, $\Pi_\ell(\bx)$, is very
different from that of a Gaussian distribution. It is spatially intermittent and
characterized by heavy tails, such that $\Pi_\ell(\bx)$ is small in
magnitude almost everywhere in space with only a few spatial regions
having very large magnitude (either positive or negative). The net (or
spatially averaged) amount of energy cascading across scales,
$\langle\Pi_\ell\rangle$ is further reduced as 
a consequence of major cancellations between upscale and
downscale transfer (positive and negative values of
$\Pi_\ell(\bx)$), accentuating the disparity between average and
extreme values. In Figures \ref{fig:SGSavgLVL01}-\ref{fig:SGSavgLVL19-26} we observe a similar tendency
for $\Pi_\ell(\bx)$ in oceanic flow. Note that the color map we use in
the figures, which has units of W/km$^2$/m, is not linear. Most of the
color shown on the map has small values close to zero (white), while some
blue/red regions exceed the maximum values on the color bar. If we
were to use a linear color map, we would register white almost
everywhere with only a few patches of blue/red in the Gulf Stream. It
is therefore important to bear in mind, when visually inspecting the
maps in Figures \ref{fig:SGSavgLVL01}-\ref{fig:SGSavgLVL19-26}, that
similar shades of red/blue may have considerably different values.

In Figure \ref{fig:SGSavgLVL01}, we see that the North Equatorial
Current, which is a shallow surface current moving westward and, in
our simulation, is at approximately 5$^\circ$N, exhibits an upscale
energy transfer. On the other hand, in Figure \ref{fig:SGSavgLVL10},
we notice that the Equatorial Counter Current, which is a deeper
eastward moving current and, in our simulation, is at approximately
100 m depth and along the equator, undergoes a downscale transfer of
energy, perhaps not unexpectedly since quasi-geostrophic dynamics is
not valid at the equator. At 500 m depth, we notice from Figures \ref{fig:SGSFlux_LVL19} and
\ref{fig:SGSavgLVL19-26} that scale-transfer is relatively weak with the
exception of the Gulf Stream which is known for its deep
penetration. This depth is well below the mixed layer within the
thermocline, where stratification effects are strongest in the ocean.

While flow in the upper ocean, above approximately 1,500 m, is mostly
due to wind forcing, the circulation in the deep ocean is mostly due
to buoyancy forcing and the meridional overturning circulation
(\citep{Talleyetal11}). The major western
boundary currents, such as the Gulf Stream, are among the more
strongly barotropic features and can penetrate all the
way to the ocean bottom.  Figures \ref{fig:SGSFlux_LVL26} and
\ref{fig:SGSavgLVL19-26} show that the relatively unstratified deep ocean
is quiescent, with energy transfer approximately $O(10^2)$ smaller
than in the mixed layer. The main activity is along the western
boundary where the flow is strongest.

\section{Comparison to Other Techniques} \label{sec:PreviousStudies}

As mentioned in the introduction, there have been several studies
published in the literature exploring the transfer of energy between
scales (\citep{ScottWang05,Tullochetal11,Arbicetal13}). These
have used somewhat different tools from the turbulence literature than have we (such as
the HT definition of flux) and used Fourier transforms to decompose scales
in wavenumber space. In this section, we explore some of the
differences between these approaches and our own.

\subsection{Tapering and detrending}
Using data from either satellites or simulations, the aforementioned
studies considered box regions in the ocean which are away from
continental boundaries. Since these box domains do not satisfy
periodic boundary conditions, the data has to be adjusted before
applying FFTs. One standard method in signal processing is 
periodizing the domain, such that the box is reflected eight times 
around the original, resulting in a ``super-box'' that is periodic
\citep{Tullochetal11}. Another standard method is to smoothly 
taper the data to zero near the edges of the box, such that the data 
becomes de facto periodic \citep{ScottWang05,Arbicetal13}.  Both of these methods can introduce artificial
gradients, length-scales, spurious acceleration and flow features not present in the original
data, although in some circumstances these effects may be small. Here, we consider only the more widely used method of tapering.

Previous studies relied on the geostrophic velocity obtained from sea-surface height (SSH) anomalies, $\eta$. These are related by
\be (u_x,u_y) = (-\partial_y \eta, \partial_x \eta ) g/f,
\ee
where $g$ is gravity and $f$ is the local Coriolis frequency. 
To illustrate the effect of tapering in the simplest possible
situation, consider a constant stream function, $\psi = \const$,
corresponding to a velocity that is identically zero. Tapering $\psi$
would introduce artificial vorticity and spurious length-scales 
which are absent in the original (zero)
flow. It should be noted that \citet{ScottWang05,Arbicetal13}
detrended $\psi$ by removing the mean and linear components of $\psi$
before tapering, such that this illustrative example does not apply to those studies.

\begin{figure}
\centering
	\includegraphics[trim=3cm 0.3cm 2cm 0.3cm,clip,width=0.8\textwidth,height=0.25\textheight]{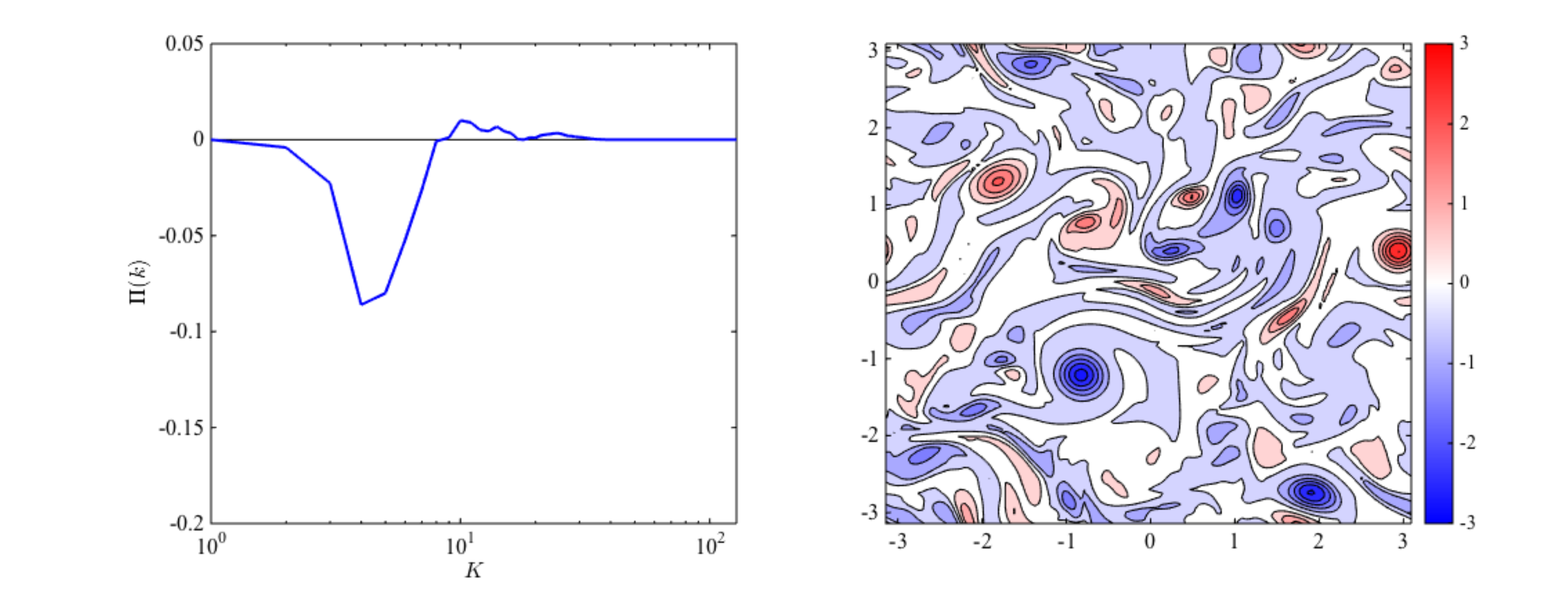}
	\includegraphics[trim=3cm 0.3cm 2cm 0.3cm,clip,width=0.8\textwidth,height=0.25\textheight]{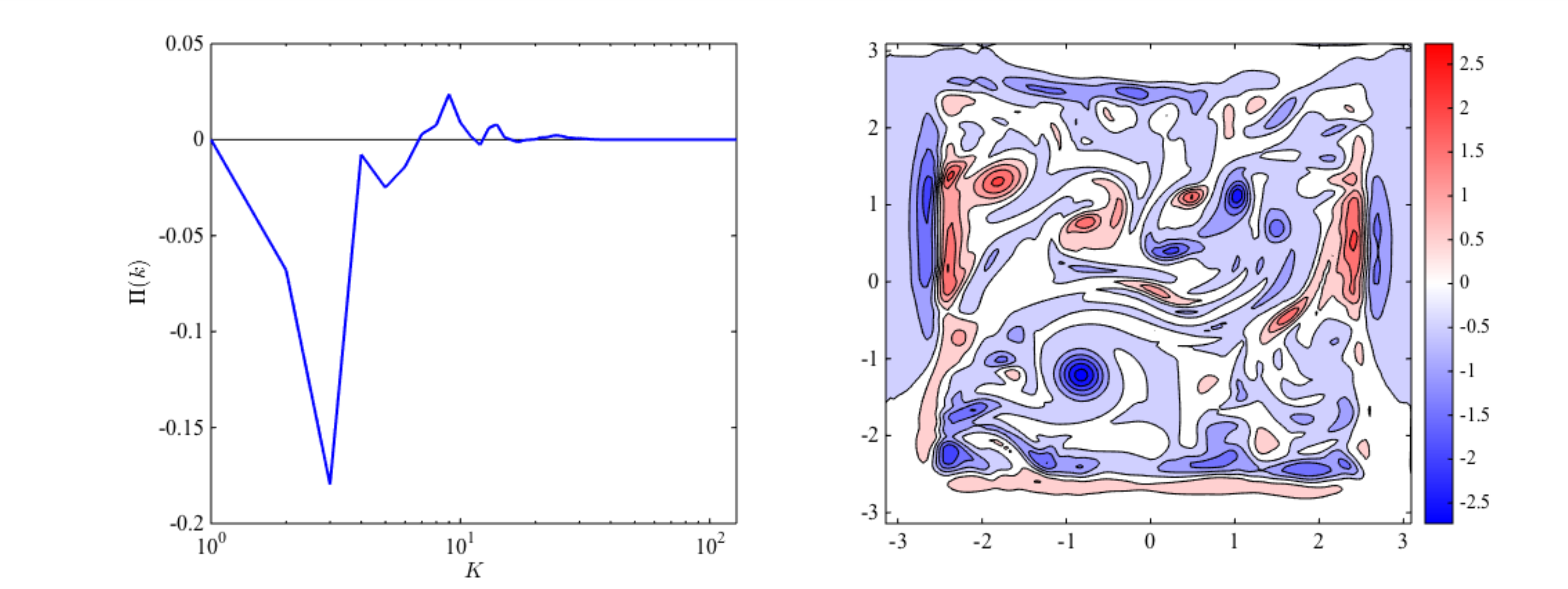}
	\includegraphics[trim=3cm 0.3cm 2cm 0.3cm,clip,width=0.8\textwidth,height=0.25\textheight]{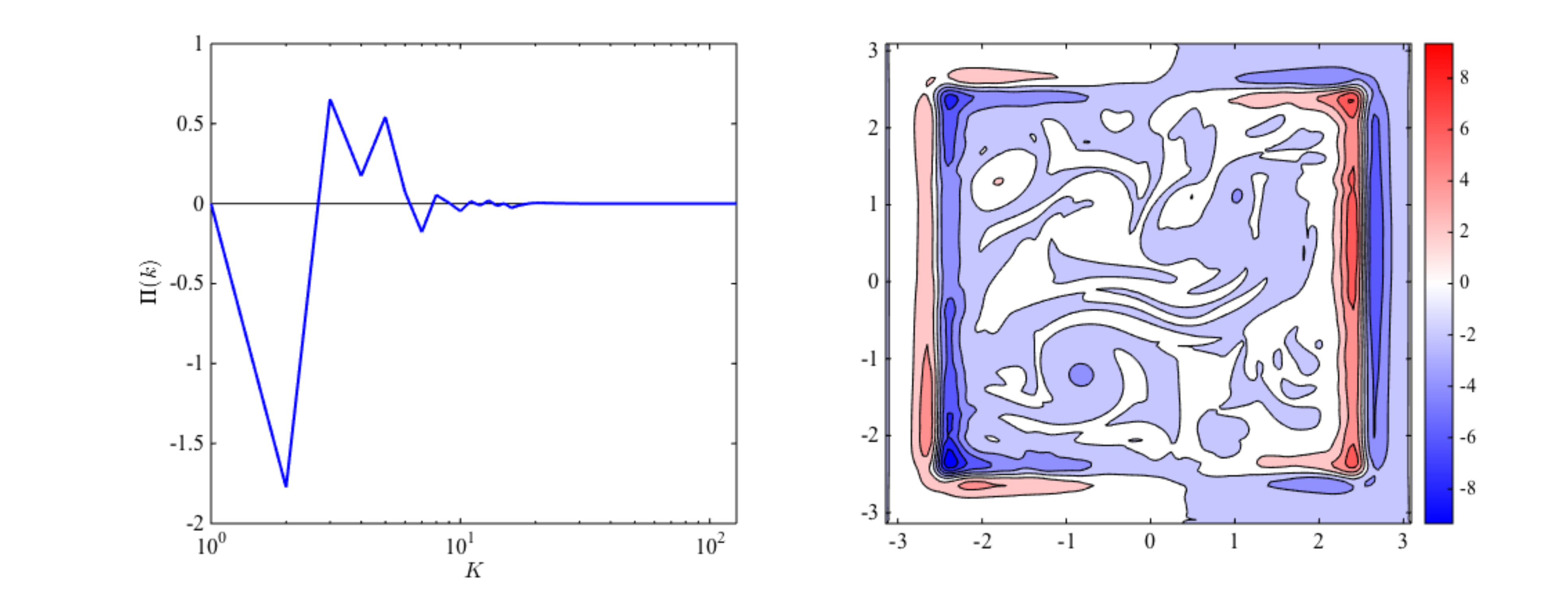}
	\caption{Left column shows $\langle\Pi_K\rangle$ as a function of scale $K=2\pi/\ell$. Right column shows vorticity contours. Top panels use original, $\psi_{{\scriptsize\mbox{orig}}}$, from a 2D flow with periodic boundaries. Middle panels use tapered streamfunction, similar to what was done in \citet{Arbicetal13}. Bottom panels adds a uniform velocity, $U_0\hat{\bf i}$, to the original flow before tapering the streamfunction, $\psi_{{\scriptsize\mbox{orig}}} + U_0y$, which illustrates the effect of tapering a jet. Even in the absence of a jet (middle), we see that tapering introduces artificial vorticity and shear to the flow, along with spurious length-scales. This is reflected as a shift in $\langle\Pi_K\rangle$ to larger scales (smaller $K$), along with significant changes in the magnitude of energy transfer.
	}
	\label{fig:TaperingEffect_Flux}
\end{figure}

However, detrending cannot remove all spurious tapering
artifacts. This is illustrated in Fig. \ref{fig:TaperingEffect_Flux},
which shows a 2D flow in a periodic box and computes the ``true''
spectral energy flux (top panel of Fig. \ref{fig:TaperingEffect_Flux})
in the simulation. In a periodic flow, the
HT and SFS flux definitions agree since the flow is
homogeneous. The middle panel of Fig. \ref{fig:TaperingEffect_Flux}
shows the flow after detrending and tapering with a window that mimics a Tukey window \footnote{\vspace{-.8cm}\be\mbox{\hspace{-1cm}Our window \hspace{1cm}} W(x,y)=W(x)\times W(y),\hspace{1cm} 
W(x)=\frac{1}{2}-\frac{1}{2}\left(\tanh\left(\frac{|x|-0.8\pi}{0.2}\right)\right),  \hspace{1cm}  (x,y)\in [-\pi,\pi)^2
\nonumber\ee
corresponds to a Tukey window with $\alpha=0.4$ and yields a value of 1 over $\approx 2/3$-rd  of the domain in each direction and tapers smoothly to zero over the remaining $1/3$-rd of the domain. This is similar to what was used in previous studies (e.g. \citep{Arbicetal13}).} (e.g. \citep{ThomsonEmery01}).
 It is clear that despite
detrending, spurious flow features and length-scales are introduced
solely due to tapering. These artifacts are also reflected in the
computed spectral energy flux, which nearly doubles in magnitude and is shifted to
larger scales due to the artificial introduction of large-scale flow
structures. Even if the flow is homogeneous to begin with, as is the
case of this periodic flow, tapering can have a substantial effect. We observe
these artifacts (shifts in scale and alteration of the flux magnitude) due to tapering 
using several windows (Hann, Tukey, and Tanh) and with different sharpness of the
tapering function. While it is common in signal processing, including in physical oceanography (e.g. Ch. 5.6.6 in \citep{ThomsonEmery01}), to compensate for any reduction in energy due to windowing by multiplying the Fourier amplitudes by an empirically determined factor, the practice only works (to some extent) for spectra but not for spectral fluxes. This is because the flux is a nonlinear quantity, which corresponds to a convolution in Fourier space. Its value at a certain wavenumber $k$ is determined by modes $k-p$ and $p$ for \emph{all} wavenumbers $p$. Therefore, a loss in amplitude at a certain mode can affect the flux at \emph{all} modes $k$ in a non-trivial way. The flux's response to window functions, unlike the spectrum's response, cannot be determined {\it a priori} because it requires knowledge of the phase relations between different modes and not just their amplitudes.

Moreover, if a flow is not
detrended, tapering artifacts may become even more
pronounced. This is illustrated in the bottom panel of
Fig. \ref{fig:TaperingEffect_Flux}. It shows the tapering effect if
the fluctuations are embedded in a mean current (which is not
detrended). In this case, the spectral energy flux increases by an order of
magnitude and shifts to yet larger scales.

\subsection{Reynolds averaging approaches}
Our coarse-graining approach is also very different from ensemble-averaging
or Reynolds averaging (RANS) 
frameworks (e.g. \citep{Gnanadesikanetal05,WatermanJayne11,Klockeretal16}) 
or density-weighted averaging \citep{Young12,MaddisonMarshall13}, 
whose essential aim is to
decompose the flow into a mean and fluctuating components. 
Several studies have relied on these types of eddy-mean flow decompositions
to analyze the energy transfer between the large-scale mean flow and the `eddy'
component of the flow (e.g. \citep{Storchetal12,Chenetal14,Kangetal15,Youngsetal17}).

A difference between coarse-graining over RANS is the freedom of the former
to choose the specific spatial scales to probe 
(Figure \ref{fig:FilteringKnEn}), which allows us to generate 
energy transfer maps across any scale 
(Figures \ref{fig:SGSavgLVL01}-\ref{fig:SGSavgLVL19-26}), and to 
quantify the energy scale-transfer as a function of scale 
(Figures \ref{fig:SGSFlux_LVL01}-\ref{fig:SGSFlux_LVL26}). 
RANS frameworks, on the other hand, usually decompose the 
flow into mean and fluctuating components without control over 
spatial scales. Moreover, the RANS description of the flow is 
inherently statistical in nature whereas the coarse-graining method 
of probing the dynamics is deterministic, allowing us to describe the 
evolution of scales at every location and at every instant in time. As 
a result, we are able to generate movies of the evolution of different 
scales.

However, the two approaches are not mutually exclusive. The 
RANS framework could in principle be incorporated into the coarse-graining 
approach since time-averaging (or, more generally, 
ensemble-averaging) and spatial-filtering are operations that 
commute with each other. One may choose to ensemble or time 
average the coarse-grained dynamics, as we have done in 
Figures \ref{fig:SGSavgLVL01}-\ref{fig:SGSavgLVL19-26}. 
Alternatively, one may spatially filter the mean and/or fluctuating 
components of the flow after having performed a RANS 
decomposition.

\subsection{Difficulties with our approach\lb{sec:Difficulties}}

Although out filtering approach does allow us to produce meaningful maps of energy scale-transfer in physical space like those in 
Figures \ref{fig:SGSavgLVL01}-\ref{fig:SGSavgLVL19-26}, it is not immune from complications. One of these (and that is also a problem with spectral approaches) is that regions close to continental
boundaries, and therefore boundary currents, require a choice to be made of boundary treatment.
If we were to filter the flow at a location adjacent to land in a standard way the filtering kernel would overlap land points, because in order to obtain the 
coarse-grained velocity, $\OL\bu_\ell(\bx)$, which is the flow at 
location $\bx$ solely composed of scales larger than $\ell$, we 
need to perform a weighted average of the velocity within a region 
of radius $\ell/2$ around $\bx$, which might include land. 
A practical choice made in this work is to treat land as water with zero velocity.
The diagnostics are then insensitive  to whether we treat land points as a solid or as water with an 
imposed zero velocity, which is consistent with the formulation of OGCMs where land is often treated as a region of zero velocity.

Another trade-off made by using our approach is due to the uncertainty principle, which prevents the simultaneous localization of a kernel in x-space and in k-space. If we were to use a kernel that is a delta function in x-space, then we are not decomposing scales ---its Fourier transform is unity, and by multiplying the Fourier transform of the velocity field with the Fourier transform of the kernel, we do not eliminate any modes (a convolution becomes a multiplication in Fourier space). The dual of such a statement is using a kernel that picks out a single Fourier mode, i.e. a kernel that is a delta function in k-space centered at mode $k_0$. Then we lose localization in x-space since the inverse Fourier transform of such a delta function gives a kernel that is a cosine wave of infinite extent in x-space. Therefore, by using a kernel that allows a certain degree of localization in x-space, we forfeit exact localization in k-space afforded by Fourier eigenmodes. These trade-offs due to the uncertainty principle are fundamental to harmonic (or spectral) analysis and, therefore, the trade-off between spatial and spectral localization cannot be eliminated. However, in our opinion, losing localization in k-space is not necessarily detrimental in situations where performing Fourier transforms is not possible, such as in the oceanic setting. Further discussion of these and related matters can be found in standard mathematics references on harmonic analysis (e.g. \citep{SteinWeiss,Krantz,Strichartz03,Sogge}). 

A practical consequence of forfeiting exact spatial localization is that oceanic 
boundaries become ``fuzzy'' due to coarse-graining. This implies, for example, that coarse-grained
velocity $\OL\bu_\ell$ can be nonzero within a distance $\ell/2$ beyond the continental boundary over
land. Therefore, terms in the large-scale energy budget (\ref{largeKE}), such as $\Pi_\ell$ and $\bJ_\ell^{\mbox{\scriptsize {transport}}}$, are only guaranteed to be zero over land a distance $\ell/2$ beyond the boundary. 
While this aspect of the method may seem undesirable, it is worth bearing in mind that such an effect also occurs in  simulations of flow over a coarse grid of cell-size $\Delta x=\ell$. 
The alternative choice is to make the filter kernel change shape as it approaches the boundary, either by making it smaller or making it conform to the boundary, but such a filtering operation will no longer commute with spatial derivatives. As a consequence of this alternative choice of boundary treatment, the coarse-graining operation would no longer preserve the fundamental physical properties of the flow, such as its incompressibility and the vorticity present at various scales. This would prevent us from deriving the large-scale energy budget (\ref{largeKE}), as discussed in section \ref{sec:FilteringScalars} above. In order to preserve these fundamental properties of the flow after coarse graining, we leave the filter independent of its proximity to the boundary.

We also want to make the reader aware of another issue pertaining to the choice of a kernel. In this work, we have made a practical choice to use a Top-Hat kernel, eq. (\ref{app_eq:Tophat_x}) above, which has a normalized value of 1 over a circular region of radius $\ell/2$ and zero beyond. In our opinion, this kernel makes the notion of scale more straightforward since it has a well-defined extent in x-space. Due to the uncertainty principle, however, it decays slowly in k-space and, therefore, may not be the best kernel to use when Fourier transforms are possible (see Fig. 13 in \citet{Riveraetal14}). There are many other kernels one could use, such as a Gaussian kernel which affords more localization in k-space (Fig. 13 in \citet{Riveraetal14}) but less localization in x-space, or the `Sinc' function, which affords the same localization in k-space as truncating the Fourier series, but has very poor localization in x-space and is more costly to implement and use. Such freedom in the kernel choice may be considered as providing flexibility but it may also be viewed as an arbitrariness in the scale-decomposition, which can influence the quantitative nature of the results. Many works have investigated the utility and drawbacks of different kernels (e.g. \citep{Vremanetal94,DomaradzkiCarati07b,EyinkAluie09,Riveraetal14}). For example, eq. (18) and Fig. 12 in \citet{Riveraetal14} discuss the energy and enstrophy fluxes across scales using different kernels. It is, therefore, important to keep these nuances in mind when interpreting results using coarse-graining, especially when comparing them to results from a purely spectral analysis when Fourier transforms are possible.

\section{Conclusion} \lb{sec:conclusion}
Understanding the transfer of energy across scales is of 
fundamental importance in oceanography. Standard methods 
based on Fourier analysis have provided important results, 
but are limited in their applicability to quasi-homogeneous 
regions with simple boundary conditions, 
and the techniques typically require some kind of special 
treatment at the boundaries. 
Information about scales of motion is not inherently tied to a 
Fourier mode decomposition, as is clear from using wavelet 
analysis or simply by high- or low-pass filtering in physical 
space. However, a straightforward application of such filters 
is insufficient to extract \textit{dynamical} information, such 
as the energy transfer across scales that can be revealed 
only by non-trivial use of the equations of motion in 
conjunction with the scale-decomposition. 

In this paper we have shown that a filtering technique, which 
we have generalized to use on spherical manifolds, 
can be used to infer information about the scales of motion, 
\textit{and} the energy transfer between scales, without being 
limited by assumptions of homogeneity or by the need to 
perform the analysis in a domain with simple boundaries. The 
technique involves a filter in physical space (a convolution 
with a kernel or window function), such as might be applied 
to smooth a field, but, moreover, used in such a way that 
\textit{coarse-grained equations of motion in physical space} 
can be derived and cross-scale energy transfer deduced. 
We have applied the technique, using full spherical geometry, 
to the results from a high-resolution eddying primitive 
equation model of the North Atlantic Ocean. 

Our method allows us to create geographic maps of the 
energy transfer. We find that an inverse energy transfer does not take place 
everywhere in the ocean (figures \ref{fig:SGSFlux_LVL10}-\ref{fig:SGSavgLVL01}) or even everywhere in the 
extratropical ocean.  In fact, certain regions are characterized 
by sustained \textit{downscale} energy transfer, such as at 
the sharp northward turn of the Gulf Stream at the Grand 
Banks, in the flanks of the core of the Gulf-Stream, and in 
the Equatorial Counter Current. These effects may be due to 
a local instability of the flow creating smaller scales or to 
non-geostrophic effects, and more analysis and results from 
observed flows will follow in subsequent papers.  In any case, 
with our method, we can clearly identify and locate regions 
where forward cascade from larger scales energizes the 
smaller scales, and we can measure the magnitude of that 
energy transfer.  

Despite the presence of regions of significant downscale 
transfer, we find that if we average over large enough regions, 
of order $10^3$ km in size or larger, away from the Equator, an 
upscale transfer is, in a basin-averaged sense, the dominant 
description of the energy scale-transfer process, confirming the 
importance of geostrophic processes on the meso- and large 
scales. Finally, we remark that the tool can also be applied to 
smaller-scale flows, such as the interaction between mesoscale 
eddies and gravity waves \citep{Nikurashin_etal13}, or in 
principle to microstructure measurements. The tool, however, 
has its limitations. As discussed in Section \ref{sec:PreviousStudies}, 
in order to spatially resolve the scale-dynamics, a certain degree of 
scale-localization must be forfeited due to the uncertainty principle. 

Our formalism can be applied to flow data from numerical simulations 
and also from satellite altimetry, as we hope will be demonstrated in future work. 
Another potential benefit of this method is in the promising area of
scale-aware modeling, where the grid resolution can vary in space,
thus requiring sub-filter models that are attuned to both geographic 
location and to the local grid scale. The coarse-graining approach 
provides a natural gateway to developing such scale-aware and 
space-aware parameterizations.

%
\section*{Acknowledgments}
We thank two anonymous referees for their comments, which helped improve this manuscript.
We also thank Robert Ecke, Burton Wendroff, Kirk Bryan, and Brian Arbic for valuable discussions, and Jonathan McLinn for developing the visualization macros that generated the scale-transfer maps. Financial support was provided by IGPPS at Los Alamos National Laboratory (LANL) and NSF grant OCE-1259794. HA was also supported through DOE grants DE-SC0014318, DE-NA0001944, and the LANL LDRD program through project number 20150568ER.  MH was also supported through the HiLAT project of the Regional and Global Climate Modeling program of the DOE's Office of Science, and GKV was also supported by NERC, the Marie Curie Foundation and the Royal Society (Wolfson Foundation). This research used resources of the National Energy Research Scientific Computing Center, a DOE Office of Science User Facility supported by the Office of Science of the U.S. Department of Energy under Contract No. DE-AC02-05CH11231. Requests for data can be sent to the corresponding author.

%






%
%
%

\end{document}